\documentclass[amsmath,12pt,amssymb]{revtex4}

\newcommand{\Bia}{\hat{\mathcal{B}}}

\usepackage{color}
\usepackage[dvipsnames]{xcolor}
\usepackage{natbib}

\usepackage{times}
\usepackage[]{graphicx}
\usepackage{amsmath}
\usepackage[caption=false]{subfig}
\usepackage{multirow}
\usepackage{refcount}
\usepackage{xr}
\usepackage{latexsym}

\usepackage[pdftex,colorlinks,citecolor=black,linkcolor=black]{hyperref} 

\usepackage[justification=raggedright,font=small,labelfont=bf]{caption}



\begin{document} 

\title{Finding influential nodes for integration in brain networks
  using optimal percolation theory}

\author{Gino Del Ferraro$^{1}\footnote{Equal contribution} $, Andrea Moreno$^{2}\textsuperscript{*}$, Byungjoon Min$^{1,3}$, Flaviano
  Morone$^{1}$, \'Ursula P\'erez-Ram\'irez$^{4}$,
  Laura P\'erez-Cervera$^{2}$, Lucas C. Parra$^{5}$, Andrei
  Holodny$^{6}$, Santiago Canals$^{2}$ \footnote{Corresponding
    author}, Hern\'an A. Makse$^{1}$\textsuperscript{$\dagger$} }

\affiliation{$^{1}$ Levich Institute and Physics Department, City College
  of New York, New York, NY 10031, USA \\ $^{2}$ Instituto de
  Neurociencias, CSIC and UMH, 03550 San Juan de Alicante, Spain
  \\ ${^3}$ Department of Physics, Chungbuk National University, Cheongju, Chungbuk 28644, Korea\\ ${^4}$ Center for Biomaterials and Tissue Engineering, UPV,
Valencia, Spain \\ $^{5}$ Biomedical
  Engineering, City College of New York, New York, NY 10031, USA
  \\ $^{6}$ Department of Radiology, Memorial Sloan Kettering Cancer
  Center, New York, NY 10065, USA }



\baselineskip 24pt



\begin{abstract}
\begin{center}
\textbf{\abstractname}
\end{center}
{\bf Global integration of information in the brain results from
  complex interactions of segregated brain networks.  Identifying the
  most influential neuronal populations that efficiently bind these
  networks is a fundamental problem of systems neuroscience. Here we
  apply optimal percolation theory and pharmacogenetic interventions
  {\it in-vivo} to predict and subsequently target nodes that are
  essential for global integration of a memory network in rodents. The
  theory predicts that integration in the memory network is mediated
  by a set of low-degree nodes located in the nucleus accumbens. This
  result is confirmed with pharmacogenetic inactivation of the nucleus
  accumbens, which eliminates the formation of the memory network,
  while inactivations of other brain areas leave the network
  intact. Thus, optimal percolation theory predicts essential nodes in
  brain networks. This could be used to identify targets of
  interventions to modulate brain function.}

\end{abstract}

\maketitle

\clearpage

\section{Introduction}

A fundamental question in systems neuroscience is how the brain
integrates distributed and specialized networks into a coherent
information processing system
~\cite{dehaene,sporns-integration}. Brain networks are considered
integrated when they exhibit long-range correlated activity over
distributed areas in the brain
\cite{sporns-integration,park2013,bullmore2009,heuvel,sporns2014}.
Correlation of brain activity is typically measured using functional
magnetic resonance imaging (fMRI), and the correlation structure is
often referred to as ``functional connectivity''
\cite{sporns-integration,park2013,bullmore2009,heuvel,sporns2014}.

Current network theory applied to such brain networks suggests that
integration of specialized modules in the brain is facilitated by a
set of essential nodes
\cite{sporns-integration,park2013,bullmore2009,gallos,morone2}. Perturbations in
such essential nodes are therefore expected to lead to large
disturbances in functional connectivity affecting global integration
~\cite{heuvel,sporns-integration,morone2}. A number of neurological and
psychiatric disorders have been attributed to disruption in the
functional connectivity in the brain \cite{heuvel,stam} and many of
the alterations associated with brain disorders are likely
concentrated on essential nodes \cite{6,12,13,14}.  Thus, identifying
these essential nodes is a key step towards understanding information
processing in brain circuits, and may help in the design of targeted
interventions to restore or compensate dysfunctional correlation
patterns in disease states of the brain \cite{stam}.

There are several studies that have used network
  centrality measures to identify the essential nodes in brain
  networks
  \cite{park2013,bullmore2009,morone2,albert,sporns2014,heuvel,stam,volkow,zuo,sporns2007}.
  These measures includes the hubs (nodes with many connections),
  betweenness centrality (BC) \cite{BC}, closeness centrality (CC)
  \cite{CC}, eigenvector centrality (EC) \cite{EC,lohmann2010}, the
  k-core \cite{hagmann,kc}, and collective influence centrality (CI)
  which uses optimal percolation theory \cite{mm} to identify
  essential nodes \cite{morone2},
  (see \cite{pei2,zuo} for a review).

These centrality measures can be used as a ranking to determine the
most influential nodes in brain networks and nodes with the highest
ranking are considered to be the ``essential'' nodes for integration.
While each centrality provides a different aspect of influence
\cite{zuo}, a common prediction of all measures is that when the
essential nodes are inactivated in a targeted intervention,
integration in the overall network is largely prevented
~\cite{heuvel,sporns-integration,morone2}. That is, when inactivated,
nodes with the highest rank lead to the largest damage to the
long-range correlations. Thus, the optimal centrality measure would be
the one which prevents integration of the network by inactivating the
fewest number of nodes \cite{kempe,mm}. The minimal set of nodes that
upon inactivation destroy the integration of the network is obtained
by mapping the problem to optimal percolation \cite{mm}. Finding this
minimal set of essential nodes is an NP-hard problem in general
\cite{kempe}. Yet, it can be approximately solved with an efficient
algorithm called Collective Influence (CI) assuming sparse network
connectivity \cite{mm,morone2}.

Some of the centrality measures have been studied using analytical and
numerical methods, and have been associated with different clinical
phenotypes \cite{heuvel,stam,zuo}. However, their importance for brain
integration has not been directly tested experimentally with
prospective interventions. The effects of removing a node from a
network has been studied with simulations, both for human and animal
brain networks \cite{lo2015,joyce2013,deasis2015}, but direct {\it
  in-vivo} validations are rare.  Thus, there is no well-grounded
approach to predict which nodes are essential for brain integration.

Here, we address this problem empirically in an {\em in vivo} rodent
preparation. We experimentally generate a network of long-range
functional connections between diverse brain areas. Specifically, we
induce synaptic long-term potentiation (LTP) in the rat dentate gyrus
\cite{bliss}, which results in correlated evoked fMRI activity in
brain areas that are involved during memory encoding and
consolidation. These include the hippocampus (HC), the prefrontal
cortex (PFC) and the nucleus accumbens (NAc) ~\cite{canals2}.  The key
question is this: Which are the essential nodes in this memory network
that are necessary for these long-range functional interactions to
form. We first identify the nodes that maximally disrupt the
integrated memory network by systematic inactivation of essential
nodes identified following the different centrality criteria.  We find
that centralities fall into two classes: hub-centralities (degree,
$k$-core, EC) which only identify the hubs at the stimulation site
(the hippocampus), and integrative centralities (CI and BC) which
identify ``weak nodes'', i.e.  low-degree yet highly influential nodes
for brain integration, notably, in the nucleus accumbens. Using
pharmacogenetic inactivation \cite{roth2016} we validate {\em in vivo}
the theoretical prediction, namely, that weak nodes in the shell of
the nucleus accumbens are essential for the integration into a larger
memory network.
These experimental results confirm the importance of going beyond the
direct connection of hubs and instead considering the collective
influence of nodes on network integration \cite{mm}.

\section{Results}

\subsection*{Overall approach}

Our combined experimental and modeling approach takes the following
steps: First, induce a functional network {\em in vivo} using synaptic
long-term potentiation (LTP) in the rat hippocampus. Second, model
this functional brain network as the result of pairwise interactions
in a sparse brain network. Third, identify and compare the essential
integrators using various centrality criteria based on the topology of
the brain network. Finally, inhibit the predicted essential and
non-essential nodes in the {\em in vivo} preparation and test whether
network integration is prevented only for essential nodes, as
predicted by the theory. In the following we elaborate on each of
these steps.

\subsection*{Experimentally coupling functional networks  {\em in vivo}} 

Long term-potentiation of synaptic connections is considered the
cellular basis of learning and memory \cite{bliss}.  Combined fMRI and
electrophysiological experiments have demonstrated that LTP induction
in the perforant pathway, the major entorhinal cortex input to the
dentate gyrus, causes a lasting increase of fMRI activity in distant
brain areas such as neocortical and mesolimbic sites (PFC and NAc)
\cite{canals2}. This result suggests that the impact of local synaptic
plasticity is not restricted to the synaptic relay at which it is
induced, as it is so usually studied, but can facilitate long-range
propagation of activity more broadly into a network formed by the
different activated areas in the brain. While this network formation
is known to depend on the activation of N-methyl-D-aspartate (NMDA)
receptors \cite{canals2}, the precise mechanisms and relative
importance of the different structures to its formation are not known
\cite{canals4}. Thus, this LTP paradigm represents an ideal system to
investigate the essential nodes for long-range integration.

We follow a well-characterized protocol to induce LTP (details of experiments in
Fig. \ref{fig:fig1}a-d and Supplementary Note 2) and apply high-frequency pulsed stimulation (250 Hz) of the
perforant pathway of the HC in six rats.  We apply low-frequency 
stimulation (10 Hz) before (PRE) and three hours after (POST) LTP induction,
to evoke
activity in the hippocampal formation while concurrently performing
fMRI. Low-frequency stimulation does not affect synaptic efficacy but does allow us to measure activated brain areas with fMRI (e.g. Fig. \ref{fig:fig1}c shows response to stimulation relative to baseline at $p<0.001$, corrected). We verify that synaptic potentiation is induced by the high-frequency stimulation by measuring the concomitant
electrophysiological recordings from the dentate gyrus as shown in
Fig. \ref{fig:fig1}b, e, and f.

LTP induction results in the propagation of evoked fMRI activity to a
long-range functional network beyond the site of low-frequency
stimulation (ipsilateral HC).  Activations after LTP induction (POST)
are reported in Fig. \ref{fig:fig1}g for a single animal, and in
Supplementary Fig. 1a for the average over six animals.
Compared to the baseline activation (PRE), we see enhanced bilateral
fMRI activation of the HC, and activation in frontal and prefrontal
neocortical regions (PFC), as well as the nucleus accumbens (NAc) (see
Fig. \ref{fig:fig1}h, i for group results and statistics;
  see also Supplementary Note 2). Conversely,
low-frequency stimulation of the perforant pathway before LTP
induction produces no fMRI activity in the prefrontal cortex nor in
the nucleus accumbens (Fig. \ref{fig:fig1}i).

\subsection*{Generate a brain network model}  \label{sec:brain_net}

The voxels with significant fMRI activation  (due to the low-frequency probe after LTP induction), form the nodes of the network model (see Supplementary Note 3 for details).
We focus on evoked activity as we are interested in propagating functional activity in the memory network, rather than spontaneous resting state activity, which will be discussed further below (Section~\ref{sec:RS}). 
The fMRI signal of the activated voxels is used to compute a functional
connectivity matrix, i.e. pairwise correlations between voxels, separately for each animal. To build the computational model of the functional network we proceed in two steps. First we identify the clusters of nodes associated with different brain areas, and then we determine the ``connectivity'' between nodes.

It is well established that the functional connectivity matrix
exhibits a modular structure, with modules (or clusters of nodes)
typically associated with different anatomical brain areas \cite{meunier}. To
identify these modules we follow standard procedures
\cite{bullmore2009}, namely, the functional connectivity matrix is
thresholded and a `community detection' algorithm is applied on this
binarized matrix \cite{girvan2002,newman2004,fortunato2010,gallos}. We also
register each brain to a standard anatomical atlas (Paxinos and Watson
rat brain atlas \cite{paxinos2007}). With this approach we identified
in each of the six animals three dominant clusters of nodes (voxels),
which overlap well with the anatomical location of the HC, the PFC or
the NAc (Supplementary Fig. 1b).

The conventional approach to generating a ``connectivity'' matrix in
brain networks models is to directly threshold the fMRI correlation
matrix \cite{bullmore2009}. However, correlations do not only arise
because two nodes exchange information or are directly linked, but may
arise due to common covariates. Furthermore, ``spurious connections''
may result from a small sample size of the time series used to compute
correlations. To minimize the effects of indirect covariation and
sampling noise we use a well establish statistical inference method
\cite{glasso}. This method models the observed correlations as the
result of direct pairwise interactions, and imposes a penalty to avoid
negligible interactions. By varying a penalization parameter, this
widely used approach tunes the sparsity of the network. As with the
direct thresholding of the correlation matrix
\cite{bullmore2009,gallos,fallani2017}, there are various ways to
select this penalization parameter. We are interested in the formation
of a connected brain network, where the different brain areas are
linked with each other. Mathematically, this corresponds to the
emergence of the ``giant connected component'' covering the entire
network, i.e. all the nodes are connected through a path
\cite{gallos,morone2}. We selected the penalization parameter that
results in the sparsest network which still exhibits a giant connected
component (see also Supplementary Note 3 for details).

In the following, the connections within each cluster are referred to
as \emph{intra-links}, descriptive of short-range interactions within
nodes in the same sub-network \cite{guimera}. Connections between
nodes belonging to different clusters are named \emph{inter-links}, or
\emph{weak-links} \cite{gallos}, reflecting the long-range
interactions between different sub-networks. Inter-links between the
HC, NAc and PFC bind these networks into a unified brain network as
seen in Fig. \ref{fig:fig1}j for a typical rat (inter- and intra-links
shown in orange and black, respectively) \cite{gallos,reis,morone2}.
Once the network model has been constructed, we proceed to identify
the essential nodes for integration.

\subsection*{Identifying essential integrators in the brain
  network model}\label{sec:integrators}

We define global integration as the formation of the largest connected
component of nodes in the network -- the ``giant connected component''
$G$. This is the graph that connects the largest numbers of nodes
through a path (highlighted in yellow in Fig.~\ref{fig:fig2}a; see
Supplementary Note 3).  The emergence of such a giant
component is an important concept in percolation theory, which studies
the behavior of clusters in networks as a function of a thresholding
parameter of the graph ~\cite{erdos,newman-random}.  The essential
integrators of the brain network are then the optimal set (minimal
number) of nodes that, upon inactivation, lead to a disintegration of
the giant component into smaller disconnected clusters. This is the
problem of optimal percolation, which attempt to find such a minimal
set of essential nodes or influencers \cite{mm,morone2}. Therefore, we
search for the essential nodes by systematic, numerical inactivation
of nodes predicted by optimal percolation theory, while we monitor the
size of the giant component.

Inactivation proceeds in rank-order according to different
centralities. We first apply the hub centrality and thus sort the
nodes by their degree. While the hub-centrality is not optimal, it is
interesting to see how the hubs rank in terms of network integration,
since they have been identified as central to integration in previous
studies. As it is customary in network theory
\cite{albert,erdos,newman-random,mm,morone2}, we quantify the damage
made to the integration of the brain network by measuring the size of
the largest connected component $G(q)$ after we remove a fraction $q$
of nodes, whereby nodes are removed in the order of degree from high
to low. Figure \ref{fig:fig2}c shows $G(q)$ under inactivation of a
fraction of $q$ hubs (mostly HC nodes in red). The curve indicates
that the inactivation of hubs does not propagate the damage to the
rest of the network. That is, removal of 20\% of hubs reduces the size
of $G$ by the same amount to 80\% of its original value for this
representative animal. Further, almost all the hubs are located in the
dentate gyrus of the hippocampus. The hub map averaged over six
animals which plots the density of essential hubs in the brain, that
is, those hubs that create the largest damage upon inactivation
(calculated in Supplementary Note 4) is shown in
Fig. \ref{fig:fig2}g and confirms that most of the essential hubs are
located at the site of LTP induction in the dentate gyrus. This is not
surprising since we stimulate its major input (the perforant pathway)
to induce the functional brain network. Inactivating the largest hubs
in the dentate gyrus experimentally would trivially disrupt the
network formation by directly preventing its local activation, rather
than by breaking the integration of the network. Thus, these top hubs
are trivial influencers.

To find essential nodes beyond the hubs at the HC, we follow optimal
percolation to estimate the minimal set of essential nodes
\cite{mm,morone2} by ranking the nodes according to the Collective
Influence (CI) algorithm \cite{morone2}.  We find that the ranking
following the CI centrality requires the smallest number of
inactivated nodes to break up the giant component since CI arises from
a maximization of the damage done to the giant component
\cite{mm,morone2}.  The CI centrality is defined by Equation
(2) in Supplementary Note 1 and
quantifies the influence of a node not only by its degree, but also by
the degree of nodes located in spheres of influence of size $\ell$ --
we refer to this as the sphere of influence ${\rm Ball}(i,\ell)$ of
radius $\ell$. Thus, CI can identify also low-degree nodes as
influential as long as they are surrounded by high degree nodes in
their spheres of influence

As shown in the particular animal in Fig. \ref{fig:fig2}c, the giant
component $G(q)$ quickly disintegrates when removing the top CI nodes
(mostly NAc nodes in green). This result is consistent across all six
animals (Supplementary Fig. 3). In clear contrast to the
results obtained for hub-nodes, Figure \ref{fig:fig2}c shows that the
removal of a very small fraction of top CI nodes ($\sim$7\% of the
total) is sufficient to reduce the giant component to 5\% of its
original size.  Crucially, most of the nodes in this influential set
are located in the nucleus accumbens as shown in the sequence of
network inactivation for this particular animal in
Fig. \ref{fig:fig2}d-f.  Figure \ref{fig:fig2}h shows the CI-map
averaged over six animals, indicating that nodes essential to brain
integration are located in the nucleus accumbens according to the CI
algorithm. This anatomical location is not predicted by conventional
hub centrality since nodes in the nucleus accumbens do not appear
among the top hubs (Fig. \ref{fig:fig2}g).
 
To illustrate the different network properties captured by
hubs and collective influence centralities consider
Fig. \ref{fig:fig2}b. Removing the node with the largest CI (depicted in black) results in large damage to the
giant connected component (shaded in blue). Removing the largest hub (depicted
in white) causes relatively less damage (shaded in red). Thus, the different nodes predicted by the hub and CI maps are the result
long-range influence encoded in the CI measure which is not captured
by the local measure of degree.  We note that the collective influence centrality
includes the  hub centrality as the zero order approximation when we
take a sphere of influence of zero radius, $\ell = 0$ in
Supplementary Eq. (2).
In this case, the  influence centrality of Eq. (2) measures the number of
connections of each node.  When $\ell \geq 1$, CI captures effects
emerging from the long-range structure.

The anatomical localization of essential nodes predicted by the other
centrality measures is shown in Fig. \ref{fig:fig3}.  A detailed
definition of these centrality measures is provided in the
Supplementary Note 1.  Betweenness centrality
(BC-map, Fig. \ref{fig:fig3}a) shares with the collective influence
centrality (CI-map, Fig. \ref{fig:fig2}h) a similar location of
essential nodes in the brain, showing that the most influential nodes
are located in the NAc shell. This indicates that the influential
nodes are also bridge nodes captured by the betweenness centrality.

In contrast, the nucleus accumbens does not appear with high $k$-core
centrality \cite{kc} (KC-map, Fig. \ref{fig:fig3}b), which shows a
distribution of essential nodes comparable to the hub map.  This
indicates that the nodes at the inner $k$-core of the network are
correlated with their degree as expected by its definition.  The
eigenvector centrality (EC-map, Fig. \ref{fig:fig3}c) also shows
essential nodes mainly located in the HC, as expected since the
eigenvectors of the adjacency matrix are highly localized by the hubs
as shown in \cite{martin}.  Finally, the closeness centrality (CC-map,
Fig. \ref{fig:fig3}d) shows essential nodes for integration in the HC
and in the NAc to a lesser extent.

These results unveil a pattern in which centrality measures dominated
by local degree (hubs, k-core, EC) tend to identify essential nodes in
the hubs of the hippocampus, since nodes with high degree are mostly
located in the hippocampus region. These nodes, in the present
experiment are trivially associated to the primary location of
stimulation, while centrality measurements that capture long-range
influence provide a non-trivial result highlighting the strength of
the low-degree nodes at the NAc.  The role of the nucleus accumbens,
thus, is analogous to a fundamental notion of sociology termed by
Granovetter as ``the strength of weak ties''
\cite{granovetter,gallos}, according to which a weak tie (in our case
a weak node, i.e. low-degree, in the NAc) becomes a crucial bridge (a
shortcut) between the densely knit clumps of close friends (the HC,
NAc and PFC).  The average map of these two categories is shown in
Fig. \ref{fig:fig3}e (hub centric: hub-KC-EC-CC-map) and
Fig. \ref{fig:fig3}f (weak-node centric: CI-BC-map). In the
Supplementary Note 7 we present the degree distribution
of the CI nodes, across animals, and compare it with the distribution
of the hubs. Supplementary Fig. 6 illustrates that most of the top CI nodes are
low-degree nodes.

Overall, this comprehensive network analysis indicates that the
integration among HC, NAc and PFC triggered by LTP induction, critically depends
on the NAc, and not only on the largest network hubs at the activation
site (HC), a fact that had not previously been recognized. The theory
based on weak-node centralities predicts that the NAc is strategically
located in the memory network, so that
inactivating a small number of its nodes is sufficient to have the
largest impact on the global connectivity; a falsifiable prediction that we test next.

\subsection*{Targeted inactivation in-vivo in the  real brain network} \label{sec:inactivation}

In order to test these predictions, we repeat the LTP experiment in an
additional five animals, while inhibiting the activity in the NAc region.  
The network module identified by the anatomic region in the nucleus
accumbens contains 33 nodes in a typical rat, corresponding to a
33mm$^3$ volume. This activated module includes the NAc core and shell
(which occupies approximately 10 mm$^3$ in the adult rat) as well as
other areas surrounding the NAc.  The theoretical prediction of CI
identifies the top influencer around coordinate 2.5 anterior and 1.3
mm lateral from bregma and 7.0 mm ventral from the cortex surface, in
Paxinos and Watson rat brain atlas space \cite{paxinos2007}.  This
location corresponds to a single node in the anterior half of the NAc
shell.  The pharmacogenetic intervention infects an approximate volume
of 1 mm$^3$, thus silencing a volume corresponding approximately to
one to two nodes (voxel volume) in the brain network structure, which
allows specific testing of the analytical prediction.

We use adenoassociated viruses (AAV) to direct the expression of
Designer Receptors Exclusively Activated by Designer Drugs (DREADDs)
\cite{roth2016} into the particular targeted area of the NAc shell
predicted as the top CI node. More specifically, we use the inhibitory
version Gi-DREADD (hM4Di) which, under intra-peritoneal administration
of the otherwise inert ligand clozapine-N-oxide (CNO), activates the
receptor inducing neuronal silencing and blocking the targeted high-CI
node in the NAc shell. With this experimental design, we
acquire fMRI data before and after administration of CNO, that is, in
presence or absence, respectively, of a functional high-CI node
located in the NAc shell of the network.

We favor the pharmacogenetic approach in this experiment over an
optogenetic strategy because it avoids implanting bilateral cannula
and optic fibers across frontal and/or prefrontal cortical regions
from which we collect and analyse fMRI signals. We microinject the
viruses bilaterally into the NAc and wait for 4 to 6 weeks to allow
strong expression of the construct (see Fig. \ref{fig:fig4}a-b and
Supplementary Note 8). Two
animals presented infection at neocortical regions due to leak of
viral particles during the injection procedure and are not considered
in further fMRI analysis.  Histological verification demonstrates that
viral expression is restricted to approximately a voxel in the shell
part of the NAc (Fig. \ref{fig:fig4}b). This subregional specificity
is most likely produced by the virus serotype used (AAV5) and gives us
the opportunity to selectively silence nodes in the NAc region
receiving most HC input \cite{groenewegen1991}.

Before LTP induction, we perform a control experiment to inactivate
the NAc shell. Comparing before and after CNO administration, (+) and
(-) respectively, we find a comparable fMRI response to low-frequency
stimulation in the hippocampus: Both the fMRI activation maps
(Fig. \ref{fig:fig4}e, g) and the amplitude of the fMRI signals
averaged across animals (Fig. \ref{fig:fig4}f, h) are unchanged,
demonstrating that the baseline fMRI response in the HC is not altered
by NAc shell inactivation. Therefore, the input necessary to drive the
formation of the memory network is preserved and can be used to
experimentally test the theoretical predictions.

Using the same animals, we induce LTP in the perforant pathway as
before but, this time, under inactivation of the NAc shell  ((+) CNO).  
Figure \ref{fig:fig4}i, j shows that, as predicted by the
theory, the formation of the long-range network
involving HC, PFC and NAc is completely prevented, yet
LTP induction still produces the expected potentiation of the
intra-hippocampal bilateral activation (compare Fig. \ref{fig:fig4}g, h and 
\ref{fig:fig4}i, j). Remarkably,
long-range inter-network links from the HC to the PFC are not formed
(Fig. \ref{fig:fig4}i, j), even though these sub-networks
 are not directly inactivated. 

For comparison, the result of LTP induction in animals with a fully
active NAc (animals without DREADD expression, (-) AAV) is shown in
Fig. \ref{fig:fig4}c demonstrating ipsilateral and contralateral HC
activation together with PFC and NAc in response to the perforant
pathway stimulation (Fig. \ref{fig:fig4}d).  These
results demonstrate that inactivation of the highest CI node in the
NAc shell disrupts the formation of the memory
  network by selectively blocking the formation of LTP-dependent
connections to neocortical structures, but not the local potentiation
of hippocampal synapses. 

\subsection*{Control experiments: \emph{in-vivo} inactivation of brain regions predicted to have no effect}\label{sec:control}

To further validate these results, we perform a series of {\it in-vivo} inactivation experiments targeting nodes which, based on our model predictions, should have no major effect on the long-range functional network.

We start with the inactivation of a node in the primary somatosensory cortex (S1), a brain region outside of HC-PFC-NAc functional network. Inactivation is first performed using DREADDs as before, with virus injection targeting the S1 region (Fig. \ref{fig:fig5}a, see Supplementary Note 8 for details). As shown by the activation maps and fMRI signals in Fig. \ref{fig:fig5}b-c, S1 inactivation does not prevent the LTP-induced activation of the HC-PFC-NAc network. Furthermore, in an additional group of animals we increased the strength of inactivation in S1 cortex by infusing 0.5 $\mu$L of tetrodotoxin (TTX, 100 $\mu$M) at the same stereotaxic coordinates (Fig. \ref{fig:fig5}d). TTX is a sodium channel blocker that completely blocks neuronal firing at these concentration (see Supplementary Note 9 for further details).  Still, Fig. \ref{fig:fig5}e-f demonstrate HC-PFC-NAc network formation upon LTP induction in these conditions.

Inactivation of the HC ipsilateral to the stimulation site would trivially eliminate the long-range network preventing its initial activation. We therefore tested whether inactivation of the contralateral HC nodes, identified by our model as non-essential nodes for global integration, would preserve network formation. 
As for S1 cortex, we used DREADDs (Fig. \ref{fig:fig5}g) and TTX (Fig. \ref{fig:fig5}j) in separate experiments to assure strong and wide inactivation of the contralateral HC (see Supplementary Note 8 and 9, for details). The results with both manipulations verify our model prediction by showing successful LTP-induced formation of a long-range HC-PFC-NAc network under contralateral HC inactivation (Fig. \ref{fig:fig5}h-i and \ref{fig:fig5}k-l). Note that TTX injection prevents the activation of the complete contralateral HC, involving a large number of network nodes but nonetheless, the long-range network is preserved. 

In our final control experiment we targeted the DREADD inactivation to the anterior part of the PFC (Fig. \ref{fig:fig5}m), a central part of the long-range network for which our model predicts low impact on global integration. TTX is not used for this target because the close proximity of the NAc and the diffusion of the TTX solution after injection cannot exclude direct inactivation of the NAc (and vice versa). However, the pharmacogenetic manipulation was enough to inactivate the PFC as demonstrated in the fMRI activation map and corresponding BOLD signals (Fig. \ref{fig:fig5}n-o). 
Most importantly, under PFC inactivation, LTP successfully recruits the long-range HC-NAc network. 

Between-groups statistical comparison (Fig. \ref{fig:fig6}, see caption for statistics) demonstrates that only NAc inactivation promotes the complete disintegration of the LTP-induced HC-PFC-NAc network, while PFC targeting only produces the expected inactivation of the PFC and control S1, and contralateral HC inactivations preserve the complete long-range integrated network. Overall, these results lend strong support to the predictive validity of the model and the key role of the NAc in the LTP-induced long-range functional network.

\subsection*{Network analysis of the Resting State dynamics}\label{sec:RS}

As already indicated, the formation of the HC-PFC-NAc network is contingent 
on LTP induction. Accordingly, prior to LTP induction, the low-frequency stimulus that probes network function, exclusively activates the HC, but neither 
PFC nor NAc are activated and, therefore, the relevance of these structures in the PRE-LTP condition cannot be studied during hippocampal stimulation. 

To shed light on the role of these brain areas before LTP induction we analyze resting-state fMRI data. From the fMRI signal prior to LTP, and in the absence of the low-frequency probing stimulus, we build a resting-state brain network for each of the six animals, by using the same network construction procedures as before.
We then use CI centrality to rank the nodes according to their importance 
for brain integration, as we did for the LTP-induced functional network. 
Further details on the procedure are discussed in Supplementary Note 5 
and an averaged CI-map over the six rats is shown in Supplementary Fig. 4. These findings should be compared 
with Fig. \ref{fig:fig2}h which presents the same type of results for the 
functional network induced by LTP.  

The outcome illustrate that, the nucleus accumbens does not always play an essential integrative role. On the contrary, the importance of the NAc arises here as a result of LTP induction. In contrast, during resting state dynamics, nodes with high CI are distributed among different brain areas
(see Supplementary Fig. 4). Therefore, the integrative role of the nucleus accumbens is specifically related to synaptic plasticity in the memory network.

\subsection*{Caveat on the methodology: from undirected to directed brain networks}
 
 Key to our reasoning is that integrating information of
specialized local modules into a global network is crucial for brain function. 
So far, this integration was modelled and measured as long-range 
correlated fMRI activity.  
However, these correlations do not necessarily measure direct
interactions between neural populations through fibers, the so-called
structural network. Some correlations may result from indirect
covariations that do not reflect direct communication between nodes.
To minimize effects due to this indirect covariations (i.e., high
correlations between two nodes that are indirect since they do not
come from a direct fiber structural connection between the nodes) we
use a statistical approach (glasso) \cite{glasso} which
attempts to explain the observed correlations as result of pairwise
interactions. However, this model assumes undirected (symmetric)
interactions.  Measuring information exchange, on the other hand,
needs a potentially asymmetric estimate that excludes some non-causal
correlation, e.g. Granger Causality \cite{granger}, which result in
directed (asymmetric) interactions.

To determine if our results are robust when directed interactions are
considered, we repeated the network analysis by endowing the network
with directed links.  For each pair of voxels in the HC-PFC-NAc
network, we determine connectivity as before
(Sec. \ref{sec:brain_net}) and, in addition, we measure Granger
causality to determine the direction of the link.
The final wiring of this directed network graph for each animal is  different from the wiring of the undirected network (see Sec. \ref{sec:brain_net}). Remarkably, by computing the CI centrality on these directed networks (see Sec. \ref{sec:integrators} and Supplementary Note 6 for details), the main results regarding the location of the influential nodes is preserved: most
 influential nodes are located in the nucleus accumbens and they are low-degree nodes, see Supplementary Fig. 5 in Supplementary Note 6. 
These results further strengthen our previous findings on the role of the NAc in the HC-PFC-NAc integration.

\section{Discussion}

While a fundamental role of the NAc in the meso-cortico-limbic system
has long been recognized, including for memory
\cite{lisman2005,pennartz2011,floresco2015}, our results suggest a new
role for the NAc function in this system. The NAc receives major
excitatory inputs from PFC and HC and dopaminergic inputs from the
ventral tegmental area (VTA), among others
\cite{groenewegen1991,floresco2015}. These anatomical, but also
neurophysiological and behavioural evidences
\cite{pennartz2011,floresco2015}, have favored the view of the NAc as
a downstream station in this circuit, working as a limbic-motor
interface with a role in selecting behaviorally relevant actions
\cite{mogenson1982}. Human and animal studies further indicate that in
addition to performing on-line processing for action selection, the
NAc encodes the output of the selected action (positive or negative
relative to expectation) into memory, which in turn will condition
future selections \cite{floresco2015,pennartz2011}. In this context,
however, our network analysis locates the NAc upstream in the circuit,
showing that interactions between the HC and PFC induced by LTP are
already under the control of the NAc. Being the interaction between
these two structures key for memory formation, we interpret our
results as indicative of a NAc-operated gating mechanism that couples
HC-PFC networks for the storage of new information, providing a
mechanism for updating memories to guide future behaviors. This
mechanism would fundamentally differ from, but being compatible with,
previous ideas on information flow between HC, PFC and NAc networks
\cite{Grace} in that the control here is exerted bottom-up from the
NAc. While the precise mechanism for this control switch has not been
investigated in the present work, an appealing possibility is the
regulation of neuronal excitability in the ventral tegmental area
(VTA) by projections of the NAc shell through the ventral pallidum
\cite{lisman2005}. In turn, dopamine release from VTA terminals in the
HC and neocortex would promote synaptic plasticity and facilitate
integration in a consolidated memory brain
  network. Regardless of the specific microcircuit, in this
network-driven theory, NAc computations seem to be a necessary part of
hippocampal-dependent memories.

The experimental model used in this work leverages the induction of
LTP in the dentate gyrus, which leads to a large-scale network that we
could perturb prospectively.  
The experimental finding highlights the importance of considering the entire network associated with each node.  Network hubs, defined solely by the number of direct connections, are not necessarily the the most effective at channeling information through the entire network.  This role may be reserved for essential nodes that connect different communities to each other \cite{power2013}.  The collective influence centrality use here accounts for the role of nodes in connecting different brain areas to one another \cite{mm}. Thus, this approach extends beyond the direct effects of hubs at integrating brain networks.

This result has
important implications for the numerous investigations on brain
pathology searching for critical alterations in functional
connectivity as disease diagnostic and/or prognostic biomarkers. A
combination of optimal percolation  theory and experimental test
presented here can be potentially adapted to networks that do not
depend on LTP induction for their formation, thus providing a recipe
to design intervention protocols to manipulate a wider range of brain
states.  These may include \cite{stam}: {\it (i)} transcranial
magnetic stimulation that can stimulate or deactivate focal brain
activity, {\it (ii)} assist in targeting deep brain stimulation
devices, in particular, for disorders that are thought to be the
result of network dysfunctions, and {\it (iii)} guiding brain tumor
surgery by identifying essential areas to be avoided during the
resection. The basic hypothesis is that activation/deactivation
patterns applied to the influential nodes will propagate through the
brain to impact global network dynamics. The proposed theoretical
analysis provides a possible road map on how to establish and test
such basic network hypotheses.

 To conclude, we mention that our analysis was based only
  on correlation structure of evoked fMRI. Future work could study the
  network structure and the role of node's degree in connectome data
  \cite{oh2014mesoscale,barabasi}. It would be important to compare the role of
  hubs, weak nodes, and nodes connecting different modules in
  structural brain networks with their role in functional
  networks. Such investigations, together with those presented in this
  work, are of crucial importance for diagnostic and clinical
  intervention in the brain. 
  
\newpage

\bigskip

{\bf Data availability} 

Data that support the findings of this study are publicly available and have been deposited in {\small \url{http://www-levich.engr.ccny.cuny.edu/webpage/hmakse/software-and-data/}}

\bigskip

{\bf Acknowledgements} 

This work was supported by NIH-NIBIB 1R01EB022720, NSF IIS-1515022, NIH-NCI U54CA137788/ U54CA132378, NSF PHY-1305476 and by MINECO and FEDER Grants BFU2015-64380-C2-1-R,
EU Horizon 2020 Grant No 668863 (SyBil-AA), and Spanish State Research
Agency, through the ``Severo Ochoa'' Program for Centers of Excellence
in R\&D (ref. SEV-2013-0317). \'U.P.-R. was supported by MECD Grant
FPU13/03537. The authors are grateful to B. Fern\'andez for technical
assistance and K. Roth for discussions.


{\bf Author contributions}

All authors contributed to all parts of the study.

{\bf Additional information}

Supplementary Information accompanies this paper at {\small \url{http://www.nature.com/naturecommunications}}

Competing interests: The Authors declare no Competing Interests. 

Correspondence and requests for materials should be addressed to
S.C and H.A.M.

\clearpage




\clearpage

\begin{figure}
\includegraphics[width=\textwidth]{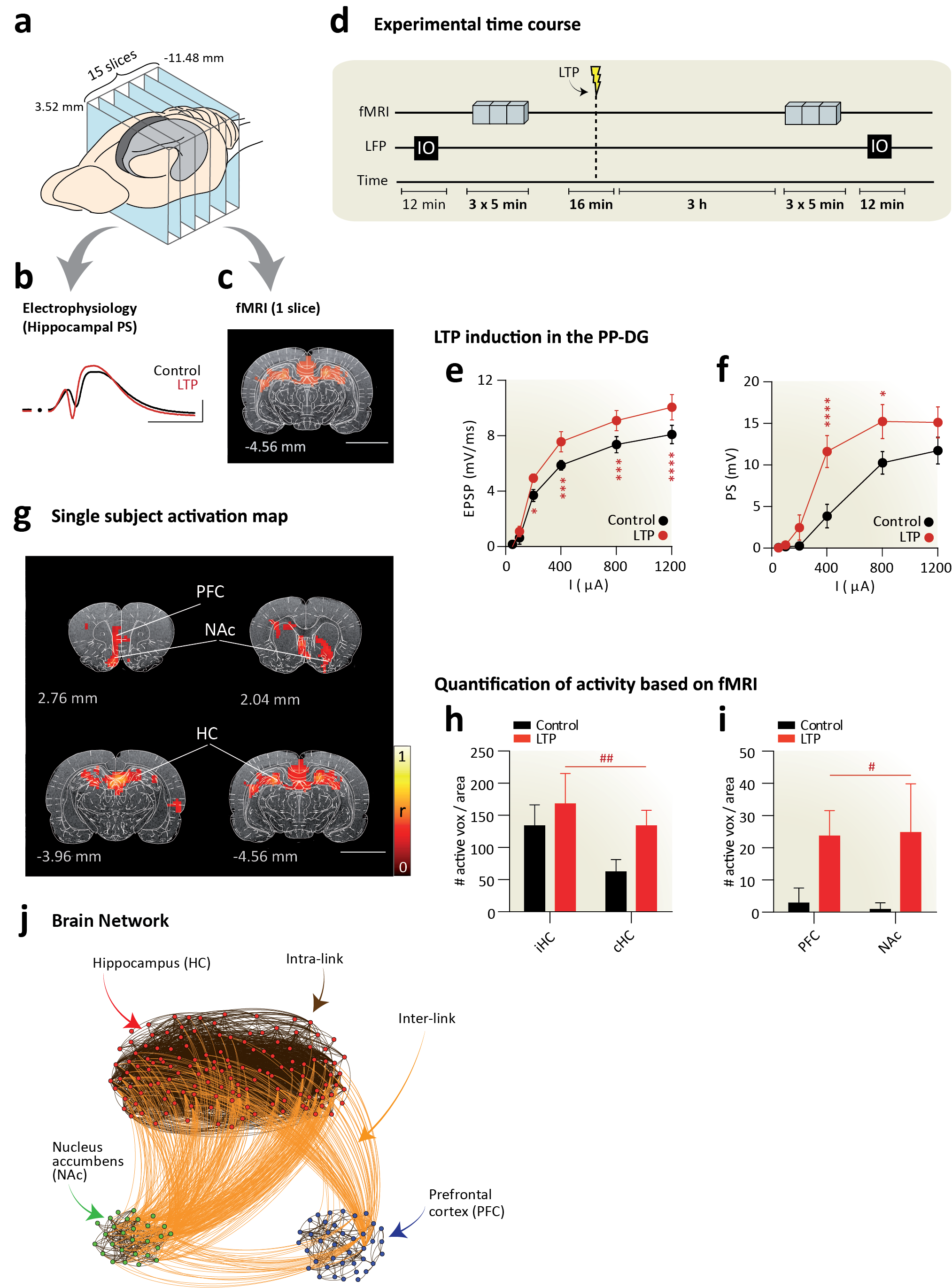}
\caption{}
\label{fig:fig1}
\end{figure}

{\bf FIG. \ref{fig:fig1}. Experimental protocol and generation of
 brain network.}   
 
{\bf a,}  Schematic representation of the imaging planes (blue). The hippocampus (HC) is highlighted in grey. Numbers indicate z coordinate in mm from bregma. 
 {\bf b,}  representative evoked population spike (PS) in the dentate gyrus before (black) and after (red) LTP induction. 
 {\bf c,} representative fMRI maps across the HC during perforant path stimulation overlaid on an anatomical T2-weighted image with atlas parcellations (see Supplementary Note 2). Color indicates significant correlation ($p < 0.005$ corrected). 
 {\bf d,} Time course of the experiment. Input/output (I/O) response curves are recorded in the local-field potentials (LFP). fMRI signals are collected during low frequency (10 Hz) test stimulations before and 3h after LTP induction. 
{\bf e,} Field excitatory postsynaptic potential (EPSP) slope and, 
{\bf f,} population spike (PS) amplitude before (black) and after (red) LTP. Two-way repeated measures ANOVA ($n=5, \alpha=0.05$) reveals significant effects of LTP in both measures ($F_{1,24}= 27.82,\, p < 0.0001$, and $F_{1,24} = 59.89; \,p < 0.0001$ for PS and EPSP, respectively). Mean $\pm$ SEM. Post-hoc Bonferroni: $^{*}\, p < 0.1; \,^{**}\, p < 0.01; \, ^{***}\, p < 0.001; \, ^{****}\, p < 0.0001$
 {\bf g,} Representative fMRI maps in one animal after LTP induction. Color code as in panel {\bf c} ($p < 0:005$; see Supplementary Note Fig. 1 for group activation maps and Supplementary Note 2 for details). Size bar corresponds to $0.5$ mm.
 {\bf h, i}, Number of active voxels per selected region in control (black) and LTP (red) conditions in hippocampal ({\bf h}) and extra-hippocampal areas ({\bf i}). The stimulated region is the ipsilateral hippocampus (iHC); two-way repeated-measures ANOVA ($n=7, \alpha =0.05$) reveals significant effects for LTP in hippocampal ($F_{1, 12} = 15.72, {\small \#\#}\, p= 0.0019$)  and extrahippocampal regions ($F_{1, 12} = 7.426,{\small \#}\, p = 0.0184$), with no interaction between regions ($F_{1, 12} = 0.00242, \,p= 0.9616$ and $F_{1, 12} = 1.518, \,p = 0.2415$ for hippocampal and extra-hippocampal regions, respectively). Mean $\pm$ SEM. 
 {\bf h}, Brain network formed by the HC, NAc and PFC for the animal in {\bf g}. The brain network is formed by intra-network interactions and inter-network interactions inferred from fMRI correlation data (Supplementary Note 3).

\begin{figure}
\includegraphics[width=\textwidth]{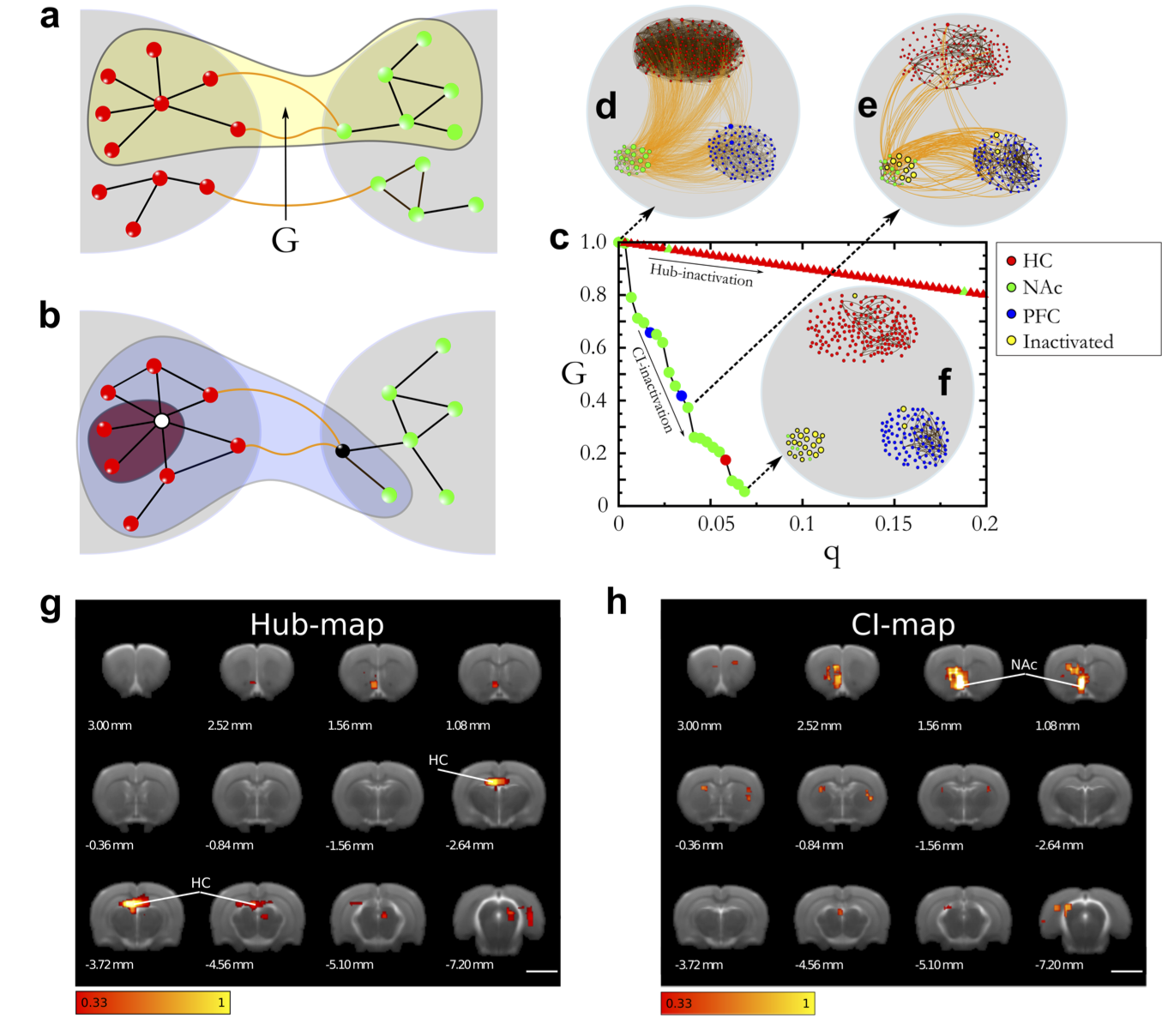}
\caption{}
\label{fig:fig2}
\end{figure}

{\bf FIG. \ref{fig:fig2}.  Hub and Collective Influence map.}  {\bf a,} The giant
(largest) connected component $G$ (yellow), captures the
integration of two modules into a brain network.
{\bf b,} Influence of a hub and a CI node.
Inactivation of the hub (white node) produces less damage to
brain  integration, measured by the
size reduction of $G$, then the inactivation of the CI node (black). 
{\bf c,} Relative size of $G$ as a function of
the fraction of inactivated nodes, $q$. Two strategies are shown for
choosing the essential nodes in a representative animal: 
Hub-inactivation (triangles) and CI-inactivation (circles). 
Nodes are removed one by one according to their degree or CI-score, respectively, from high to low.
Colors refer to the nodeÕs module (HC, NAc, or PFC, see legend). Most hubs
(red symbols) are located in HC, yet, they are not essential for
integration: their removal makes minimal damage to $G$. 
On the contrary, by inactivating 7\% of
high CI nodes, $G$ collapses to almost zero.  Most CI 
nodes are in the NAc (green symbols). 
{\bf d,} Representative brain network as in {\bf c}, displaying the
PFC-HC-NAc networks. The size of each node is proportional to
the CI-score. 
{\bf e,} We inactivate
the top 3\% of high CI nodes (yellow circles) and $G$ is drastically
reduced to less than 40\% of its original value.  These top CI nodes
are all in the NAc except for two nodes in the PFC.  
{\bf f,} Further
inactivating up to 7\% of the high CI nodes prevents
integration of $G$. Yellow circles indicate the essential nodes,
located mostly in the NAc shell.  
{\bf g,} Average hub
map indicating top hub nodes over six animals.  Yellow/white areas correspond to top essential nodes all located in the HC since this is the area of LTP
induction.  Color bar represents the average rank (Supplementary Eq. (8)).  
{\bf h,} Average CI map
indicating top CI nodes over six animals, most CI nodes results in the NAc and are
generally not hubs.  Color bar is defined in Supplementary
Eq. (8), the size bar corresponds to $0.5$ mm.

\begin{figure}
\includegraphics[width=\textwidth]{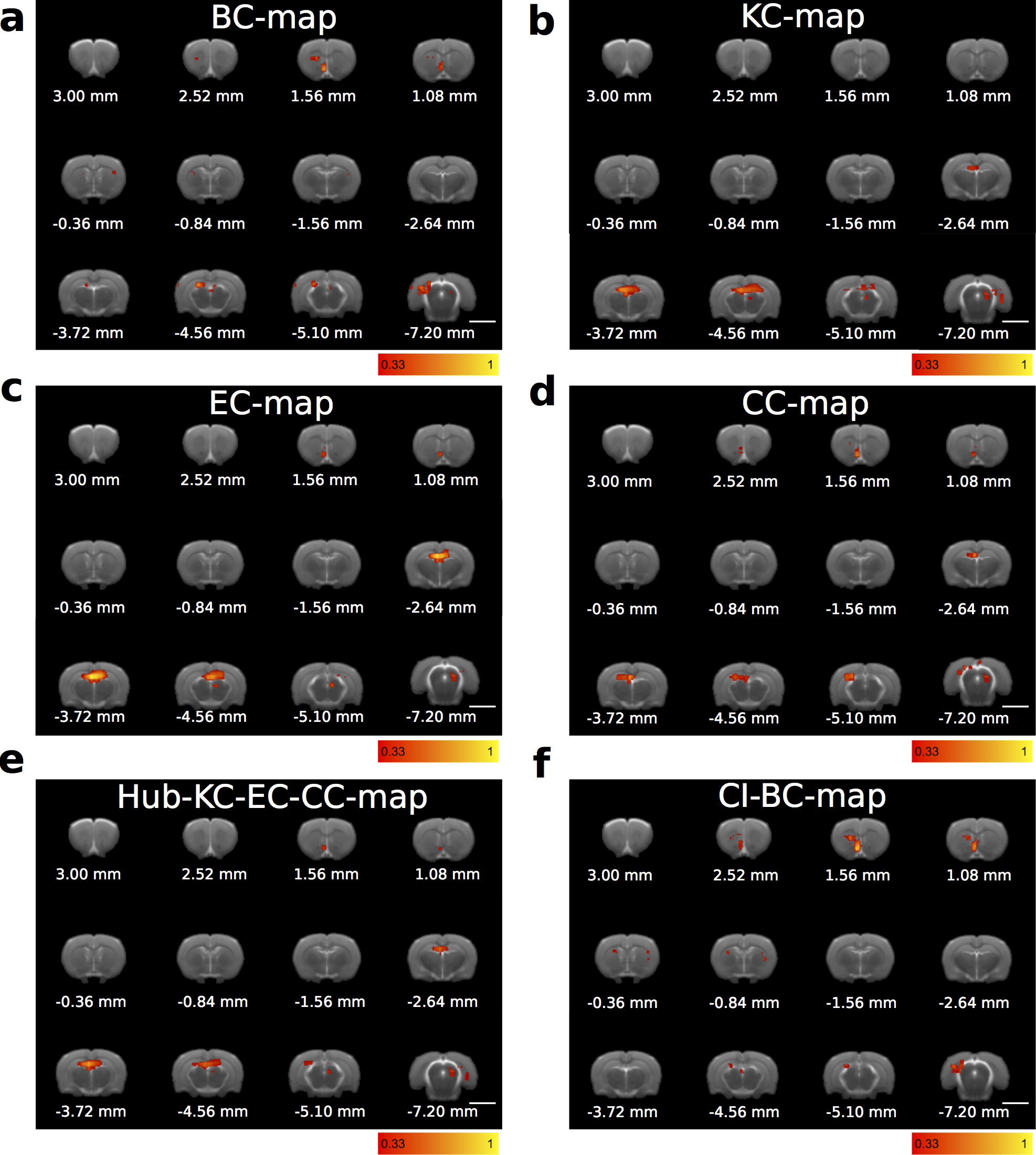}
\caption{}
\label{fig:fig3} 
\end{figure}

{\bf FIG. \ref{fig:fig3}. Maps of essential nodes.}  Average map of
influencers for the different centralities according to {\bf a,}
betweenness centrality, {\bf b,} $k$-core centrality, {\bf c,}
eigenvector centrality and {\bf d,} closeness centrality.  The maps
are average over the six rats and the color bar are calculated
according to the rank defined by Supplementary Note 4, Eq. (8).
Yellow/white colors indicate the top influencers according to each
centrality. According to these results, the centralities are then
divided into {\bf e,} hub-centric centralities dominated by the hubs
and identifying the hubs in the HC and {\bf f,} integrative
centralities dominated by the weak nodes and identifying the low
degree nodes in the shell part of the NAc. The size bar in each panel
corresponds to $0.5$ mm.

\begin{figure}
\includegraphics[width=\textwidth]{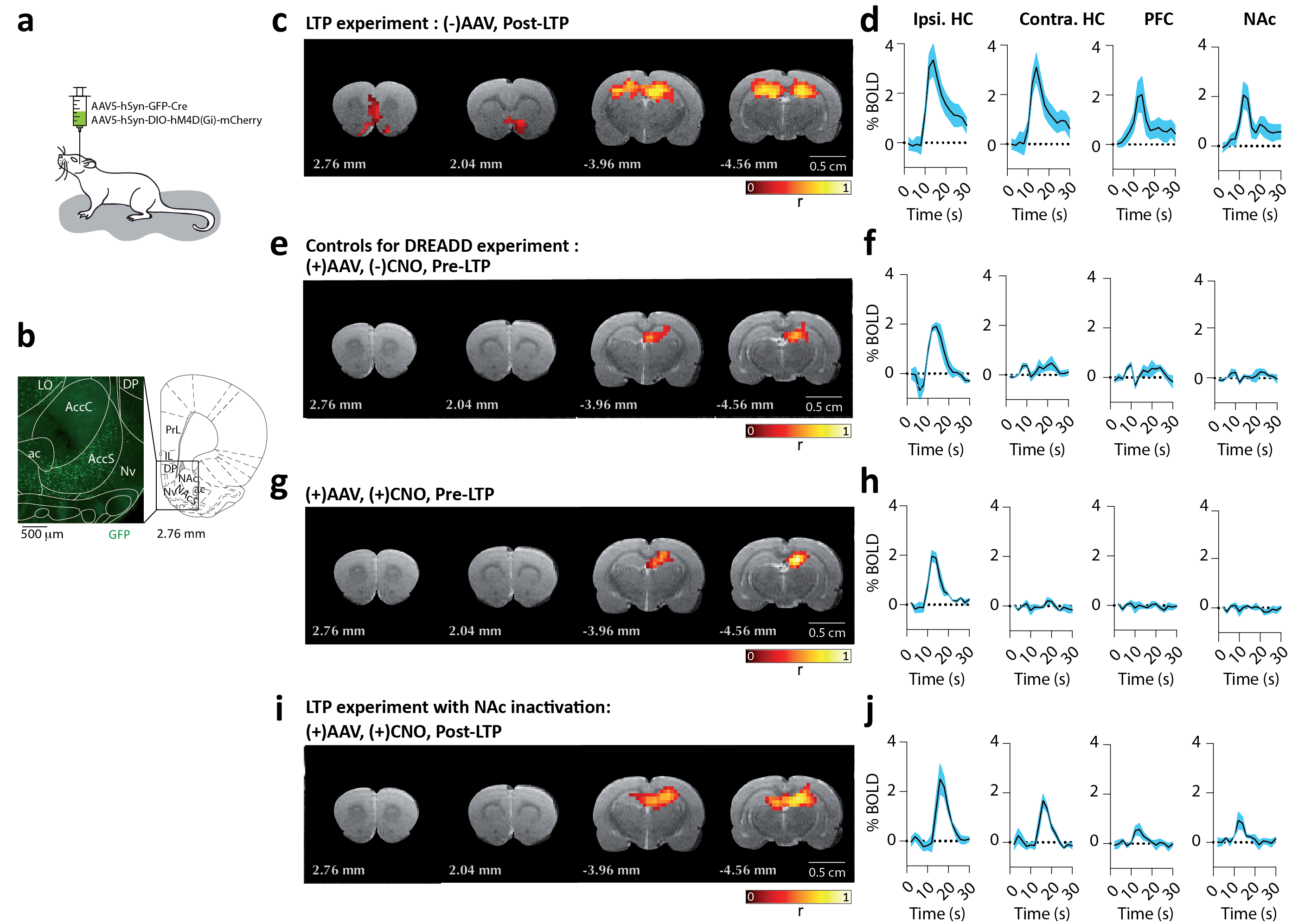}
\caption{}
\label{fig:fig4}
\end{figure}

{\bf FIG. \ref{fig:fig4}. Experimental test of essential nodes.} 
{\bf a,} The inhibitory version of DREADD receptors (hM4Di) is expressed in the NAc shell using a combination of two adenoassociated viruses (AAVs) injected stereotactically in the region as indicated (see Supplementary Note 8 for details). 
 {\bf b,} Histological verification 4 weeks after the viral infection showing green fluorescence protein (GFP) in the neurons that positively express the construct. For anatomical reference, an image of the rat brain atlas is shown. Inset: 20x magnification picture of the same slice demonstrating selective infection of neurons in the NAc shell.
{\bf c, e, g, i:} single subject statistically thresholded fMRI maps showing voxels activated ($p < 0.001$, corrected) by perforant path stimulation and overlaid on an anatomical T2-weighted image. {\bf  d, f, h, j:} BOLD time courses from significantly activated voxels averaged from the indicated regions of interests and across animals (mean $\pm$ SEM; $n = 6$ for panel {\bf c}, $n = 3$ for panels {\bf e, g} and {\bf i}). Details on fMRI processing and statistics are given in Supplementary Note 2 and 8. 
{\bf c,} LTP experiment for comparison ((-) AAV infection, (-) CNO administration) showing the expected activation of HC, PFC and NAc in POST-LTP. Note the evoked BOLD responses bilaterally in the HC (panel {\bf d}), a landmark of HC response after LTP induction. 
{\bf e,} AAV infection in the NAc ((+) AAV, (-) CNO) preserves activation of the HC under perforant path stimulation before LTP. {\bf g,} Inactivating the NAc by administration of CNO in the same animal ((+) AAV, (+) CNO) does not alter functional maps nor BOLD responses in the baseline (PRE-LTP) condition (compare {\bf e} vs {\bf g}). BOLD signal responses ({\bf f, h}) are only evident in the ipsilateral HC as expected from PRE-LTP condition.
{\bf i, j,} NAc inactivation ((+) AAV, (+) CNO) prevents the integration of the long-range network involving HC-PFC-NAc induced by LTP (POST-LTP).

\begin{figure}
\includegraphics[width=\textwidth]{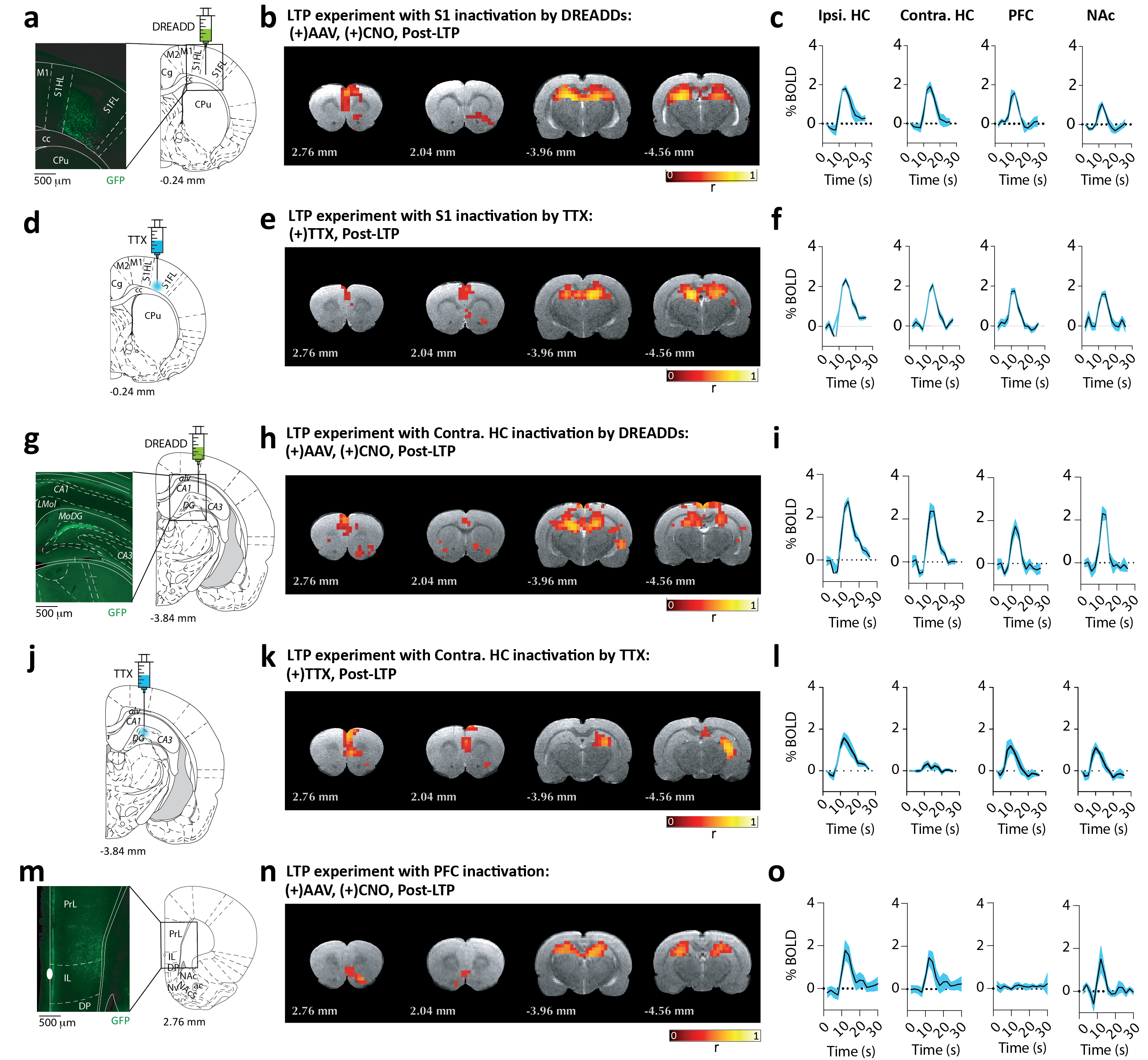}
\caption{}
\label{fig:fig5}
\end{figure}

{\bf  FIG. \ref{fig:fig5}. Targeted inactivation of different brain regions}. 
{\bf a, b, c,} Pharmacogenetic inactivation of S1 cortex. {\bf a,} location of the AAVs injection in the corresponding section of the stereotaxic map and representative histological staining showing the construct expression (inset). {\bf b,} shows the statistically thresholded ($p<0.001$, corrected) fMRI maps of a representative animal and {\bf c,} the averaged BOLD signals across subjects ($n=2$) and across region of interest. As in Fig. \ref{fig:fig4}, S1 inactivation does not disrupt the long-range network formed upon LTP induction.
{\bf d, e, f,} Inactivation of S1 with TTX (See Supplementary Note 9, for experimental details). Same as a-b-c experiments with S1 inactivation using the sodium channel blocker TTX ($n=3$). Both fMRI maps and BOLD signals demonstrate formation of the HC-PFC-NAc network triggered by LTP ($p<0.001$).
{\bf g-l,} Pharmacogenetic (g, h, i, $n= 5$) and TTX (j, k, l, $n=4$) inactivation the contralateral HC ($n=5$). As shown in the individual fMRI maps and averaged BOLD signals ($p<0.001$), none of the inactivation strategies targeting the contralateral HC prevented the formation of the HC-PFC-NAc network.
{\bf m, n, o,} Pharmacogenetic inactivation of the PFC ($n=5$). AAVs injection targeted to the anterior part of the PFC ({\bf m}) prevents its activation by performant path stimulation, as expected by the pharmacogenetic intervention, but does not abolish the formation of the long-range HC-NAc connections ($p<0.001$), as predicted by the theory ({\bf n, o}).


{\bf FIG. \ref{fig:fig6}.  Group analysis of network inactivation.}
The number of nodes in the relevant networks is quantified after LTP induction with or without targeted inactivation and normalized relative to the control, fully active, condition.
{\bf a,} Proportion of nodes recruited by perforant pathway stimulation in the ipsilateral HC under control conditions (100\%) and after inactivation of the NAc, PFC, contralateral HC (cHC) and S1, as indicated in the $x$-axis. Analysis of variance across groups demonstrates no statistical differences (ANOVA, $F_{4,24}=0.3641, p=0.8317$).
{\bf b,} Proportion of nodes recruited in the NAc following targeted inactivation in the structures indicated in the $x$-axis. ANOVA demonstrates statistically significant differences between groups ($F_{4,24}=4.841, p=0.0053$) and post-hoc Bonferroni test finds the only significant difference in NAc recruitment when the NAc is directly inactivated ($p<0.01$), as expected from the experimental manipulation, but no effect under PFC, cHC or S1 inactivation.
{\bf c,} Proportion of nodes recruited in the PFC following targeted inactivation in the structures indicated in the $x$-axis. ANOVA demonstrates statistically significant differences between groups ($F_{4,24} = 6.416, p=0.0012$) and post-hoc Bonferroni test identifies strong reductions in both PFC ($p<0.01$), expected from the experimental manipulation, but also NAc ($p<0.05$), indicating the disintegration of the long-range HC-PFC-NAc network under NAc inactivation.


\clearpage

\clearpage


\clearpage


\clearpage


\clearpage


\begin{figure}
\includegraphics[width=\textwidth]{Figure_6_2018_rev.ai}
\caption{}
\label{fig:fig6}
\end{figure}

\clearpage





\centerline{\bf {\Large Supplementary Information}}

\vspace{1cm}


\centerline {\bf {\large Finding influential nodes for integration in
    brain networks using optimal percolation theory}}
 
 \bigskip
 
 Del Ferraro et al.




\setcounter{section}{0}
\setcounter{figure}{0}

\renewcommand{\figurename}{Supplementary Figure}

\section{Finding the essential nodes for integration in the brain network}
\label{si-centrality}


In this section, we provide the heuristic algorithms used to identify
influential nodes. For each algorithm, we assign the score to each
node by following the described algorithms and sort the nodes
according to the score.


{\bf Degree centrality}. Degree centrality is the number of nearest
neighbors in the network. Degree centrality is one of the simplest
metric for identifying important nodes.  Hubs refer to nodes in the
network with large degree.

{\bf $k$-core and $k$-shell index} \cite{seidman,kc,hagmann}.
$k$-core (KC) refers to a subset of nodes formed by iteratively
removing all nodes that have degree less than $k$. In other words,
$k$-core is a maximal subgraph where all nodes have at least $k$
neighbors. $k$-shell index is then the largest $k$ value of $k$-core
that the node belongs to. To assign $k$-shell index for each node, we
first delete all nodes with degree $k=1$, iteratively. The removed
nodes via the process belong to $k$-shell with $k_S=1$. We remove next
$k$-shell with $k_S=2$ and we proceed to remove all the higher shells
iteratively until all nodes are removed. Then, we can assign a unique
$k$-shell index to each node in networks. It has been shown that the
importance of hub nodes can be highly diminished if they are located
in the periphery of the network, i.e., the low $k_s$ shells. On the
other hands, nodes in the inner $k_s$ shells define the core of the
network and correspond to the influencers in the network
\cite{kc}. However, by its own definition, the nodes in the inner
shells are generally high degree nodes, therefore the $k$-core
centrality is highly correlated with the degree.

{\bf Collective Influence} \cite{mm,morone2}.  Collective influence
(CI) is designed to approximately identify the minimal set of nodes
that can produce disconnected networks, based on optimal percolation
network theory \cite{mm}.  Mathematically, the problem can be mapped
to optimal percolation and can be solved by the minimization of the
largest eigenvalue of the non-backtracking matrix of the network
\cite{mm,morone2}.  This optimization theory was originally developed
for single networks in \cite{mm} and was extended to the case of brain networks
in \cite{morone2} in the context of brain network of networks.
The activation of nodes in the brain network was described by a state
variable $\sigma_i$, which acts as an ON and OFF switch ($1$ and $0$,
respectively) to reflect the activation/inactivation state of node
$i$.  If a node is directly inactivated, then $\sigma_i=0$.  A node
can also be inactivated indirectly as a result of lacking input from
its inactivated neighbors in the other network, which, mathematically,
is equivalent to the McCulloch-Pitts model of neuronal activation
\cite{mcp}:
\begin{equation}
\begin{aligned}
\sigma_i\ &=\ 0\ \ \ &{\rm direct\ inactivation}\ ,\\ 
\sigma_i\ &=\ \Theta \Bigg(
\sum_{j\in\mathcal{N}(i)} \sigma_j \Bigg) \ \ \ &{\rm otherwise} .
\label{eq:sigma2}
\end{aligned}
\end{equation} 
The sum in the second equation reflects the integration of incoming
activity from all nodes $j$ that connect to node $i$ from other
networks $\mathcal{N}(i)$, and the threshold operation via the
Heaviside step function $\Theta$ indicates that a minimum of incoming
activity is needed for activity to propagate \cite{mcp}.

The collective influence (CI) score assigned to each node $i$ in the
 brain network in this model is given by \cite{morone2}:
\begin{equation}
  {\rm CI}_{\ell}(i) = (k_i-1)
\sum_{j\in\partial{\rm Ball}(i,\ell)} (k_j-1) +
  \sum_{ \substack{j\in\mathcal{F}(i)\ : \\ k_j^{\rm inter}\ = 1}} (k_j-1)   
\sum_{m\in\partial{\rm Ball}(j,\ell)} (k_m-1) .
\label{eq:CIeq}
\end{equation} 
Here, $k_i\ \equiv\ k_i^{\rm intra}+k_i^{\rm inter}$ is the degree,
$k_i^{\rm intra}$ is the number of connections of node $i$ within its
network, $k_i^{\rm inter}$ is the number of connections to nodes in
different networks in the set $\mathcal{F}(i)$, and $\partial{\rm
  Ball}(i,\ell)$ indicates the sphere of influence of node $i$ at
distance $\ell$.

Technically, CI is the contribution of each node to the eigenvalue of
the non-backtracking matrix, which determines the stability of the
giant component \cite{mm}. CI is an optimization measure that attempts
to find the smallest set of nodes that will produce the largest damage
to the giant connected component of the brain network, which is analogous to
minimize the largest eigenvalue of the non-backtracking matrix
\cite{mm} defined on the $2M \times 2M$ edges of the network (in the
case of single networks):

\begin{equation}
    \mathcal{B}_{k\to \ell, i\to j} =
\begin{cases}
1,  \,\,\,\,\mbox{if} \,\, j=k \,\,\,\, \mbox{and}\,\,\,  i\neq \ell\ ,\\
0,  \,\,\,\, \mbox{otherwise}.
\end{cases}
\label{non}
\end{equation}

Thus, the matrix $\mathcal{B}_{k\to \ell, i\to j}$ has non-zero
entries only when $(k\to \ell, i\to j)$ form a pair of consecutive
non-backtracking directed edges, i.e. $(k\to \ell, \ell\to j)$ with
$k\neq j$. In this case $\mathcal{B}_{k\to \ell, \ell\to j}=1$. The
powers of the matrix $\Bia$ give the number of non-backtracking walks
of a given length between two nodes in the network
\cite{hashimoto,nbt}, in analogy to the powers of the adjacency matrix
which count the number of paths \cite{newman-book}.


The CI algorithm runs as follows \cite{morone2}: i) at the beginning,
we choose the value $\ell$ of the radius of the Collective Influence
sphere.  In our analysis of the brain network, we use the value
$\ell=2$.  We find that higher values of $\ell$ give nearly the same
results since the networks contain short paths. The value of $\ell$ is
always smaller than the largest path in the network, and it can be
optimally chosen by systematically changing it from $\ell=1$ to the
diameter of the network. We find that the optimal set of nodes is
obtained when $\ell=2$.  ii) Next, CI for all nodes is computed using
Eq.~\eqref{eq:CIeq}, and the node with the largest CI is
inactivated. iii) Then, the CI values of the remaining active nodes
are recalculated, and the next highest CI node is inactivated. iv)
Step iii) is repeated until the giant active component vanishes.

{\bf Betweenness centrality} \cite{BC}.  Betweenness centrality (BC)
measures the influence of nodes based on the shortest paths on
networks. BC for each node is defined as the number of the shortest
paths that pass through the node. BC identifies crucial nodes for
information flow and packing transportation by definition. This
centrality can capture low-degree nodes that are strategically located
between large communities. For instance, imagine a node with $k=2$
with each link connecting to a large community of tightly connected
nodes.  Such a low degree node will have a large BC since all the
paths between nodes in the two distinct communities will necessarily
pass through this bridge node.

{\bf Eigenvector centrality} \cite{EC}.
Eigenvector centrality (EC) is defined as the entry of the eigenvector 
that corresponds to the largest eigenvalue of adjacency matrix defined as

\begin{equation}
A_{ij}=
\begin{cases}
1, \,\,\,\,\mbox{if} \,\, i \,\,\,\, \mbox{and}\,\,\, j \,\,\,\,
\mbox{are connected}\\
0, \,\,\,\, \mbox{otherwise}.
\end{cases}
\end{equation}
The main idea of EC is that the influence of nodes is determined by
the importance of its neighbors. Therefore, neighbors with
high-scoring eigenvector centrality more contribute to the score of
the node. PageRank is also a variant of EC. It has been proved in \cite{martin}
that the use of the largest eigenvalue of the adjacency matrix can lead
to a localization of the influence in the hubs. Thus the EC centrality
is highly correlated with the high degree and contains similar
information about the influencers. This localization problem is solved
by replacing the adjacency matrix in the centrality by the
non-backtracking matrix Eq. (\ref{non}).

{\bf Closeness centrality} \cite{CC}.  Closeness centrality (CC) is
defined as the inverse of the average distance of shortest paths
between the node with all other nodes in the network.  The higher
closeness is, the closer it is to all other nodes in average.  In
practice, closeness play an important role in transportation since
nodes with higher CC can disseminate information efficiently to the
whole connected network via shortest paths.  This centrality is mainly
determined by the degree since hubs will naturally be closers to other
nodes in the networks, thus, it is considered as one of the
hub-centric centralities.

\section{{Experimental Design and Long-term potentiation experiments}} 
\label{protocol}

The brain network is based on long-term potentiation (LTP)
experiments. LTP is a synaptic strength modification protocol that
leads to changes in neuronal networks, and is believed to be one of
the key mechanisms by which the brain undergoes memory processes
(acquisition, consolidation, and extinction)
\cite{bliss,lynch,bliss1973,bliss1993,martin2000,morris2001,squire1986}.
It refers to the enhancement of synaptic transmission efficacy in
specific neuronal connections. This mechanism has been observed to
occur under natural learning conditions, yet, experimental
manipulation of synaptic transmission has allowed deciphering many of
its characteristics, dissecting the synaptic plasticity process from
other on-going processes during memory formation. In the present work,
we use experimental LTP induction in the rat hippocampus to provide an
experimental model of controlled long-range functional connectivity
reorganization.

All experiments were approved by the Spanish authorities (IN-CSIC),
CCNY Institutional Animal Care and Use Committee Review of Research
Protocol No. 980, and were performed in accordance with Spanish (law
32/2007) and European regulations (EU directive 86/609, EU decree
2001-486). The data used in this study can be found at:
{\small \url{http://kcorelab.org}}.  Details of the experiments are explained
in the next sections.

\subsection*{Subjects}

A total of 37 Sprague-Dawley male rats, weighing between 250-350 g,
were used in these experiments. From these, 29 animals were conserved
 for data analysis (6 controls for the LTP network generation in baseline conditions, 
 4 for NAc inactivation with DREADDs,  5 for PFC inactivation with DREADDs, 
 4 for Hippocampal inactivation with DREADDs, 5 for Hippocampal inactivation with TTX, 
 2 for S1 inactivation with DREADDs, and 3 for S1 inactivation with TTX. 
 A total of five animals were discarded due to surgery complications or
poor quality of MR images, and additional three because of leak of viral
particles to the neocortex in the NAc inactivation
experiments. Animals were purchased from Janvier Labs (France) and
maintained under a 12/12 h light/dark cycle (lights on 07:00-19:00 h)
at room temperature (22$\pm$2 C). Food and water were provided ad
libitum. Rats were housed in groups (4-5 animals per cage) and adapted
to these conditions for at least 7 days before any manipulation.

\subsection*{Surgery and electrode implantation}

The animals are anesthetized briefly with isoflurane (3-4 $\%$ isoflurane 
in 0.8 L/min O2 flow) and then injected intraperitoneally with 
urethane (1.3 g/kg). After 60 minutes, the main reflexes disappearance is 
tested and, if necessary, a second dose of urethane is injected (1/5 of 
the initial dose) as reinforcement. When reflexes disappear the surgery 
starts. During the complete procedure animals are maintained with constant 
temperature (37.0-37.5 C) with a water pad. Vital constants (pulse and 
breath distension, heart and breath rate, and oxygen saturation) are 
monitored using a paw-clip pulse oximeter (MouseOx Plus, Starr Life 
Sciences, Oakmont, US). A constant flow of $O_2$ (0.8 L/min) is supplied 
through a mask.

The anesthetized animal is placed in a stereotaxic frame (Narishige,
Japan) and a local anesthetic is injected subcutaneously in the
incision points (0.2 mL of bupivacaine). The skin is opened and
retracted with suture thread hold to haemostat clamps to expose the
bone surface. Special care is taken to remove all traces of blood from
the skull and mussel that would decrease MRI data quality due to
susceptibility artefacts. Care during surgery is maximized to prevent
even minor spontaneous bleeding throughout the MRI session which would
also distort the BOLD (blood oxygenation level dependent)
signal. Trephine holes are made by hand with a manual driller (2 mm
diameter) in the target coordinates and the dura is pinched with a
curved needle at the incision points to allow the penetration of the
electrodes.

A bipolar stimulation electrode made of twisted platinum-iridium 
wires (Teflon coated, 0.025 mm diameter, WPI, USA) is inserted in the 
perforant pathway, a bundle of axonal fibers that represents the principal 
input of information to the hippocampus (AP 0.0 mm from lambda; ML 4.1 mm 
from lambda; DV 2.1-2.5 mm from brain surface). A recording multichannel 
electrode (multichannel recording electrode, 32 channels, model 
A1x32-6mm-100-177, NeuroNexus, Ann Arbor, Michigan, USA) is lowered in 
the ipsilateral dorsal hippocampus (AP 3.5 mm from bregma, ML 2.5 mm from 
bregma, DV 3.5 mm from brain surface). Electrophysiological recordings 
are made in order to precisely position the stimulating electrode in its
optimal location based on the evoked potential recorded in the hippocampus. 
Once in place, the multichannel recording electrode is replaced by a 
single channel recording probe (MRI compatible) in the dentate gyrus of 
the ipsilateral dorsal hippocampus. Both stimulation and recording electrodes 
are implanted in the brain with acrylic dental cement 
(SuperBond, Sun Medical, Japan) and bone cement (Palacos, Heraheus Medical 
GmbH, Germany) and the animal is then transported into the MRI facility.

\subsection*{Electrophysiological recordings}

A single pulse stimulation protocol (100 $\mu$s bipolar pulse,
delivered at a 0.05 Hz rate) is recorded before and after LTP
induction to assess synaptic potentiation (Fig.
\ref{fig:fig1}a).  To this end, an Input-Output curve is obtained at
different stimulation intensities (50, 100, 200, 400, 800, 1000, and
1200 $\mu$A) while recording the evoked field potentials in the
dentate gyrus. After filtering (0.1 Hz -- 3 kHz) and amplification,
the electrophysiological signals are digitized (20 kHz acquisition
rate) and stored in a personal computer for offline processing with
Spike2. The population spike (PS) in the hilus of the DG is measured
as the amplitude from the precedent positive crest and the negative
peak, and the excitatory postsynaptic potential (EPSP) is measured as
the maximal slope of the raising potential preceding the PS.


\subsection*{fMRI measurements}
\label{fmri}

Imaging experiments are carried out in a 7 Tesla scanner with a 30 cm
bore diameter (Biospec 70/30v, Bruker Medical, Ettlingen,
Germany). Acquisition is performed in 15 coronal slices using a GE-EPI
sequence applying the following parameters: FOV= 25.25 mm; slice
thickness= 1 mm; matrix= 96 $\times$ 96; segments= 1; FA, 608; TE= 15
ms; TR =2000 ms. This provides a resolution of the raw images of
0.26$\times$0.26$\times$1 mm.

Additionally, T2 weighted anatomical images are collected using a
rapid acquisition relaxation enhanced sequence (RARE): FOV= 25.25 mm;
15 slices; slice thickness= 1 mm; matrix= 192$\times$192; TEeff= 56
ms; TR= 2 s; RARE factor= 8. A 1H rat brain receive-only phase array
coil with integrated combiner and preamplifier, and no tune/no match,
is employed in combination with the actively detuned transmit-only
resonator (Bruker BioSpin MRI GmbH, Germany).

Once in the MRI scanner, the anesthetized animal is constantly
supplied with a 0.6-0.8 l/min O2 and heated with a water-bath system
to keep a constant temperature (37 $\pm$ 0.5 C). Physiological
constants are measured as before using a paw-clip pulse oximeter
(MouseOx Plus, Starr Life Sciences, Oakmont, US) equipped with a MRI
compatible cable.  Functional MR images are acquired before (Pre
condition) and after LTP induction (POST condition) using a
low-frequency 10 Hz stimulation protocol that activates the
hippocampal formation without altering synaptic plasticity, as shown
before \cite{canals2,canals4,canals3,canals2008}.  This stimulation
consists of a block design protocol as follows (see
Fig. \ref{fig:fig1}d): ON periods lasting 4 s of 40 pulses train, each
composed of a 10 Hz stimulation train at 800 $\mu$A. We follow the ON
period by OFF period with no stimulation for 26 s. This ON/OFF
sequence is repeated 10 times, for a total of 300 s.

LTP is induced inside the MRI scanner using a high frequency
stimulation (HFS) protocol, consisting of 6 bursts of 8 pulses each
delivered at 250 Hz, with bursts repeated 6 times with a 2 minute
separation between them. The total duration of the protocol is 960
s. MR images are not acquired during LTP induction. Three hours after
induction, the same low-frequency stimulation protocol as used for the
PRE-LTP condition (10 Hz) is used and fMRI acquisition is performed to
record the consequences of synaptic potentiation on functional
connectivity.

Functional MR images are preprocessed separately using FSL 5.1 M
\cite{jenkinson2012,smith2004} and AFNI \cite{cox1996,cox2012}
tools. First, the images are converted from Bruker to NIfTI
format. Then, motion is corrected by aligning each volume to the mean
image volume \cite{jenkinson2002}, slice timing correction is applied,
and the brain is extracted \cite{smith2002}. The next step is to
obtain the transformation matrix to register the functional images to
a rat brain T2-weighted MRI template \cite{schwarz2006}.  This
registration Mark \cite{jenkinson2002,jenkinson2001} is performed in
two steps: 1) functional images are aligned to anatomical images using
a rigid-body transformation and 2) anatomical images are
affine-registered to the standard template. Both matrices are
concatenated but not applied to the functional images, which remained
in their native space. The inverse transformation is used to bring the
regions of interest (i.e hippocampus, prefrontal cortex, nucleus
accumbens and the venous sinus) from the Paxinos and Watson rat brain
atlas \cite{paxinos2007} to the functional space. The venous sinus is
removed from the images. Afterwards, spatial smoothing using a 2-mm
FWHM (full width at half maximum) Gaussian kernel is applied, followed
by mean-based intensity normalization to obtain a global 4D mean of
10,000. Subsequently, linear and quadratic trends, global signal and
six motion parameters (three translations plus three rotations) are
regressed out. Finally, the time series are bandpass temporally
filtered [0.01-0.1] Hz via Fast Fourier Transform.  After this process
a BOLD signal as a function of time, $x_i(t)$, is output for every
voxel $i$ in the brain. This signal is the basis for the construction
of the brain network model as we explain next.

\section{{Method to construct the LTP brain network}}
\label{si-non}

After the BOLD signal has been obtained for every voxel in the brain,
we construct the brain network model via the following procedure: (1)
Identification of statistically significant activated voxels
(activation map) $\to$ (2) Calculation of correlation $C_{ij}$ between
all pair of voxels in the activation map $\to$ (3) Identification of
brain modules through clustering algorithms $\to$ (4) Inference of
interactions $J_{ij}$ between pairs of voxels using graphical-lasso
$\to$ (5) Determination of essential influential nodes using the CI
algorithm from optimal percolation theory.

\subsection*{Activation map}
\label{actmap}

We first determine which brain voxels are activated by the
low-frequency stimulation protocol using the
FEAT analysis tool in FSL ({\small \url{https://fsl.fmrib.ox.ac.uk/fsl/fslwiki/FEAT}}). 
The regression assumes for the explanatory
variable the block-design of the low-frequency stimulation as
described above.  After the general linear model (GLM) analysis, 
the $Z$ statistic map is thresholded and cluster corrected 
(cluster $Z$ threshold $= 2.3$). 
Figure \ref{fig:fig1}g shows the activation map for a single animal in
the POST-LTP condition. In Supplementary Figure \ref{si:fig6}a we show
the same activation map but averaged over the six animals.  This map
represents voxels that are activated in the POST-LTP state in at least
2 out of 6 animals with $p<0.001$ (determined after co-registering the
fMRI recordings to a common anatomical rat brain atlas of Paxinos and
Watson \cite{paxinos2007}).  Supplementary Figure \ref{si:fig6}b shows
the anatomical areas corresponding to the HC, PFC, and NAc. Comparison
between both images indicates that voxels in these three areas are
activated after the LTP induction.  These activated areas form the
basic voxels used as ``nodes'' in the subsequent calculation of the
brain network model.

\begin{figure}
\includegraphics[width=\textwidth]{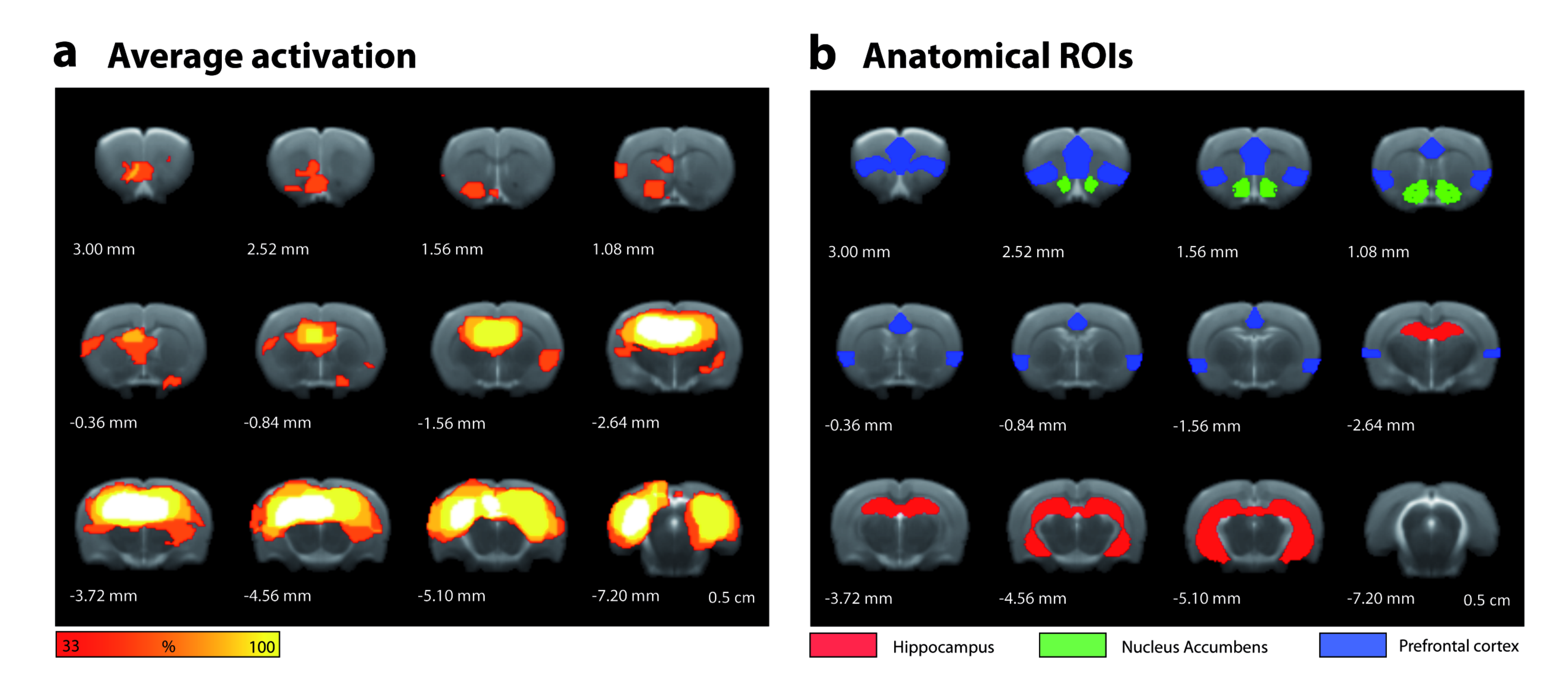}
\caption{{ Activation map and anatomical areas of interest.} {\bf
    a,} Group (n=6) average activation map after LTP induction. This
  map represents voxels that are activated three hours after LTP
  stimulation (POST condition) in at least 33\% of the animals with
  $p<0.01$ (see Supplementary Note \ref{actmap}). Note activation in the
  hippocampus (HC), prefrontal cortex (PFC) and nucleus accumbens
  (NAc). Numbers indicate distance from bregma.
{\bf b,} Anatomical map defining the three main areas of study: HC,
PFC and NAc.}
\label{si:fig6}
\end{figure}

\subsection*{Construction of memory networks}
\label{construction1}

In order to construct the brain network we first compute the
correlation coefficients or sample covariance $C_{ij}$ of the BOLD
signal between voxels $i$ and $j$ in the activation map, often referred
to as ``functional connectivity'':
\begin{equation}
C_{ij}=\frac{\langle x_i x_j \rangle - \langle x_i \rangle \langle x_j \rangle}{ \sqrt{(\langle x_i^2 \rangle -\langle x_i \rangle^2) (\langle x_j^2 \rangle -\langle x_j \rangle^2) } },
\label{Cij}
\end{equation}
where $x_i(t)$ is the BOLD signal of voxel $i$ as a function of time
$t$ and $\langle \cdot \rangle$ represents the temporal average over
the recording period.  Correlations are computed separately for each
animal for all voxels that showed significant activation in at least 2
animals (activation maps were co-registered to a standard atlas, but
correlation is computed in the original space to avoid introducing
spurious correlations due to resampling).

In the animal original space, the BOLD signal is measured at a
resolution of 0.26$\times$0.26$\times$1 mm.  Another source of
spurious correlations might arise when applying the customary spatial
smoothing to the image with a Gaussian kernel, because the volume
space is not isotropic. So, to avoid including spurious correlations
of fMRI signals in the $(x,y)$-plane, we consider only every four
voxels so that nodes are separated by $1.04\times1.04\times1$ mm, and
are approximately isotropic in all three dimensions.  Therefore, the
size of the voxel, that is, each node in the brain network, is
approximately 1 mm$^3$ and this corresponds to a single node in the
network. This size is commensurate with the size of the target in the
pharmacogenetic interventions.  The same downsampling procedure
described above is applied in all the analysis described in the text,
with or without pharmacogenetic intervention.
Following existing literature we model these correlations as the
result of pairwise interactions between nodes
\cite{park2013,robinson,robinson2,deco,sarkar}.

\subsection*{Inference of the connections of sparse network} 
\label{inference1}

The pair-wise correlation modelling
literature typically assumes that brain networks have sparse
connectivity \cite{bullmore,deco,robinson,robinson2,sarkar}. We
therefore construct sparse graphs by using machine learning techniques
like the graphical Lasso algorithm \cite{glasso}. Given normal
distributed data, the log-likelihood for observing the sample
covariance ${\bf C}=\{ C_{ij} \}$, defined in Eq. (\ref{Cij}, is given
by the log of the Wishart distribution:
\begin{equation}
\log L({\bf J}) = \log \det({\bf J}) - {\rm Tr}({\bf C}{\bf J}),
\end{equation}
where ${\bf J}=\{ J_{ij} \}$ is the model for the inverse
covariance. These $J_{ij}$ reflects the strength of interactions
between a pair of nodes $i$ and $j$.  To implement the assumption of
sparse interactions the Graphical Lasso algorithm assumes a Laplace
prior, which results in a maximum a posteriori estimate with a L1-norm
penalty term \cite{glasso}:
\begin{equation}
{\bf J}^*=\underset{{\bf J}}{\rm argmin} [{\rm Tr}({\bf C}  {\bf J}) - 
\log\det({\bf J}) + \lambda |{\bf J}|],
\label{eq:glasso}
\end{equation}
where $ |{\bf J}|$ is the L1-norm of the interaction matrix and
$\lambda$ is the penalty parameter controlling how sparse the
estimated ${\bf J}^*$ will be. A sparse interaction matrix will have
many zero entries. A non-zero entry indicates that there is a
pair-wise interaction, while $J_{ij}^*=0$ means that there is no
direct interaction between $i$ and $j$. We infer the sparse
matrix $J_{ij}$ fixing the $\lambda$ penalization parameter in Eq. 
\eqref{eq:glasso} as described below, for each separate animal.

Since we are interested to study the integration of a set of networks
aggregated into a giant connected component, we define the brain
network via a procedure involving a change in the penalty parameter
$\lambda$, which tunes the sparsity of the network (see
Eq. \eqref{eq:glasso}).  A giant component is a connected component of
a given graph that contains a constant fraction of the entire graph's
vertices in the thermodynamic of an infinite system size.  As
$\lambda$ is changed from a high value to a low value, a series of
networks emerge to form the giant connected component of brain network
in a procedure that we explain below.  Higher values of $\lambda$
penalize almost all of the links and therefore the brain network is
disconnected.  As we reduce the values of $\lambda$ in
Eq. \eqref{eq:glasso}, more links appear and the brain network
transforms into a giant connected components of nodes (inside this
component there is a path connecting every pair of nodes). For a
finite graph, we consider the giant component as the largest connected
component in the graph and study the behaviour of its relative size
$G_{\rm bond}$ as a function of $\lambda$.  In these plots, $G_{\rm
  bond}$ represents the ratio of nodes belonging to the largest
connected component to the total number of nodes in the brain
network. The suffix bond refers to the fact that this process builds
the brain network via a process analogous to bond percolation (see
below) \cite{bollobas,erdos,bollobas2,percolation}.  Thus, we use the
'bond' denomination of this giant connected component since it is
constructed by adding links to the network by reducing the penalty
parameter $\lambda$. Indeed this process is analogous to
bond-percolation and attempts to solve the problem of choosing the
thresholding or penalty parameter that defines the binary network from
the weighted covariance matrix by using the concept of the emergence
of the giant connected component. That is, following \cite{gallos} we
choose the penalty in such a way that the resulting network is at the
point of emergence of the connected components that connect each
cluster HC, PFC and NAc in turn. This process results into a sparse,
yet, connected network and it follows the idea that the most important
feature of the network that we want to capture in our study is the
long-range connectivity and integration of the different components
into a unitary network.  Thus, we whole analytical procedure starts by
findings the sparse connected network of HC-PFC-NAc via
graphical-lasso and bond percolation of the penalty parameter to then
apply the optimal percolation method via the collective influence
algorithm to find the essential nodes for inactivation.  We explain
this procedure next.

\begin{figure}[!t]
\includegraphics[width=8.5cm]{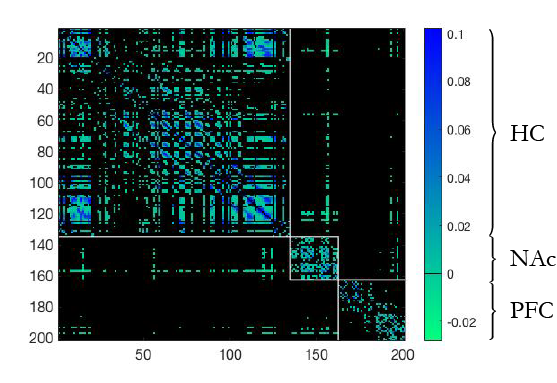}
\caption{For the same representative animal of Fig. \ref{fig:fig1}g
  and \ref{fig:fig1}j: Adjacency matrix of the resulting
  brain network, obtained by bond-percolation using the penalty
  parameter $\lambda$ in the graphical lasso algorithm as described in
 Supplementary Note \ref{construction1}. Nodes
  are ordered according to their membership to one of the anatomical
  clusters: HC, PFC and NAc. From above to below, the first module
  corresponds to the HC, the second to the PFC, the third to the NAc.}
\label{fig:J}
\end{figure}

In a percolation problem one monitors the size of (fraction of nodes
belonging to) the giant connected component $G_{\rm bond}$ as a
function of the driving external parameter.  In the present case, we
first apply the graphical lasso for a given $\lambda$ and obtain the
inferred matrix $J_{ij}$ from $C_{ij}$. We binarize this matrix and
construct a network by considering a link if $|J_{ij}|$ is above a
given small resolution threshold, as it is customary in the graphical
lasso algorithm.  We then monitor the giant component of this network
for a given $\lambda$ versus the penalty parameter $\lambda$ and we
search for the appearance of the giant component as $\lambda$ is
decreased from a large value. The process of constructing the network
by decreasing $\lambda$ adds links to an initially empty network as in
bond percolation. We fix the penalty parameter $\lambda$, which tunes
the sparsity of the network, as the highest value at which the giant
component of the network appears between each cluster, in turn.
i.e. such that all nodes in the three clusters HC-PFC-NAc are
connected through a path. In other words, the final network is the
sparsest architecture that yet has one connected giant component which
includes nodes from the three clusters. The connectivity matrix is
obtained by binarizing the obtained $J_{ij}$ from the graphical lasso
at a given $\lambda$ by considering a link when $J_{ij}$ is non-zero
with a given small resolution. The resulting connectivity matrix from
$J_{ij}$ is shown in Supplementary Fig. \ref{fig:J}, for the same
representative animal used in Fig. \ref{fig:fig1}g and
\ref{fig:fig1}j.  From this matrix we
identify the three anatomical components HC, PFC and NAc and the links
inside the clusters or strong or intra-links and the links across the
clusters, the weak or inter-links \cite{gallos}.

\begin{figure}
\includegraphics[width=0.8\textwidth]{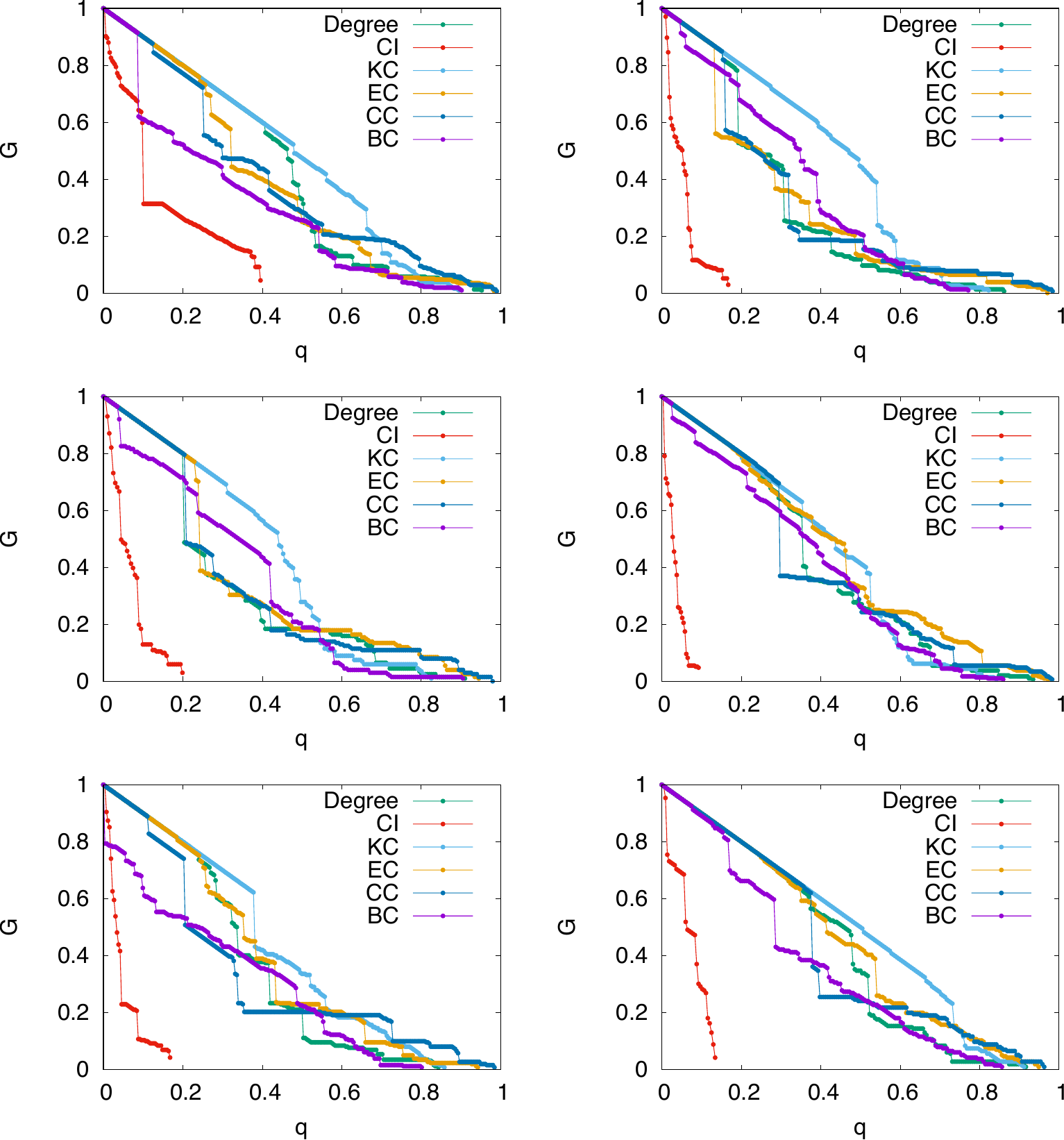}
\caption{
  Size of the giant connected component $G$ as a function of
	the fraction of inactivated nodes, $q$ for all six rats
	for degree, CI, KC, EC, BC, and CC. For CI, smaller 
	number of inactivated nodes are required to disintegrate the network
	consistently for all six rats. 
}
\label{si:fig8}
\end{figure}

\section{{Measure of average maps of centralities in the
  brain network }}

\label{maps}

For all nodes in the brain network we compute the score of each
centrality for each experimental animal. We then rank all the nodes
from high to low score.  We then 'attack' the brain network following
each ranking for each centrality from hubs, CI, KC, EC, CC and BC. We
monitor the size of the giant component as we remove a fraction of
influential nodes $q$ following each strategy and for each network
corresponding to each of the six animals \cite{mm,morone2}. Supplementary Figure
\ref{si:fig8} shows the results. We see how the strategy following CI
destroys the giant component with the smallest number of nodes. For
each strategy, we extract the set of most influential nodes, the
essential nodes according to each strategy, by considering the first
nodes that reduce the size of the connected component to 5\% of its
original size. These are the set of essential nodes for each
centrality and correspond to the ranking of top nodes according to
each centrality.

Lastly, we normalize the ranking of each node using the following
formula \cite{morone2} to compare across strategies:
\begin{equation}
R(i) = \frac{r_o-r_i}{r_o-1},
\label{SI:eq-CI-norm}
\end{equation}
where $r_i$ is the ranking of node $i$, that is defined as the step at
which it is inactivated (for example, the first node to be inactivated
is assigned $r_i=1$, the second $r_i=2$, and so on). The quantity
$r_o$ is a baseline, which, in our analysis, we set as the ranking of
the node for which the giant active component takes the value
$G=0.05$. Note that $R(i) = 1$ represents the highest node.  On the
other hand, if node $i$ is not targeted by an external inactivation,
then we set $R(i) = 0$.  The normalization in
Eq.~\eqref{SI:eq-CI-norm} allows us to properly sum over all samples
to get an averaged map of the most important nodes in the brain
network which allows us to compare the impact of each centrality. The
results are used to generate the the hub-map in Fig. \ref{fig:fig2}g
and the averaged CI map in Fig. \ref{fig:fig2}h, as well as all the
centrality maps shown in Fig.  \ref{fig:fig3}.

\section{{Influencers map for the resting state dynamics}}\label{sec:SI-RS}

In this section we present results regarding which nodes are
responsible for integration during resting state dynamics, as
discussed in Sec. \ref{sec:RS}. The analysis of the essential nodes
for integration presented in the main text, indeed, is performed on
brain networks stimulated by LTP induction which, in addition to the
hippocampus, produces the activation of the prefrontal cortex and of
the nucleus accumbens. In the PRE-LTP condition, stimulation of the
hippocampus does not recruit activation of neither the PFC nor the NAc
and therefore, the relevance of these latter areas for brain
integration cannot be investigated. To clarify their role in the brain
network, we analyze the fMRI signal of the resting state dynamics in a
PRE-LTP condition.

Since we are interested in investigating the role of the HC, the PFC and the NAc during unperturbed brain dynamics, we take into consideration the same anatomical areas, i.e. same voxels, studied to analyze the LTP-induced network. This guarantees that nodes in the resulting brain networks are the same for both the LTP-induced and the resting-state network. What changes between the two cases are the BOLD signals and, therefore, the statistical dependences between these voxels, i.e. the wiring of the resulting architecture.

\begin{figure}[t!]
\includegraphics[width=7.8cm]{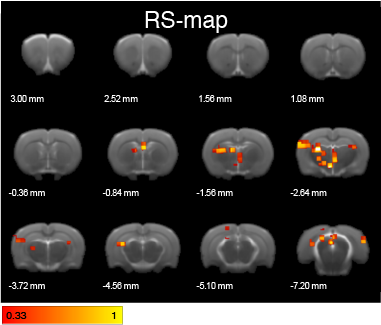}
\caption{ Average ($n=6$) CI-map for the resting-state brain
  dynamics. The map shows the CI-score (Eq. \eqref{eq:CIeq}) of each
  voxel averaged over six animals for the case of unperturbed brain
  PRE-LTP induction. High CI-score voxels are not localized in a
  single brain areas but appears spread around and mostly located
  between the hippocampus and the prefrontal cortex.}
\label{fig:RS}
\end{figure}

The analysis is done on the same six animals presented in the main text (in that case POST-LTP), same $p$-value ($p < 0.001$). Each resting-state brain network is constructed similarly to the LTP-induced one, as described in Sec. \ref{sec:brain_net} and in Supplementary Note \ref{si-non}. For each of these networks we rank the nodes according to the CI centrality measure, obtaining a CI-map for each rat. We then average the CI-score across the six animals, similarly to the LTP-induced networks, as described in Supplementary Note \ref{maps}. The averaged results are shown in Supplementary Figure \ref{fig:RS} which shows no role for the NAc as director of brain integration. High CI-score nodes, indeed, are less localized to a single brain area and are rather spread in different brain regions, mostly involving the hippocampus and the prefrontal cortex. These findings demonstrate that brain integration is related to brain dynamics. The role of the NAc as director of brain integration discussed in the main text is not simply arising because of its anatomical location in the brain but, rather, it is due to the functional re-organization stimulated by LTP-induction.

\section{{Directed brain network analysis}}\label{sec:directed}

The network analysis of influencers in the rodent brain presented in
Sec. \ref{sec:brain_net} and \ref{sec:integrators} is based on the
construction of an undirected network. All biological networks are
directional and so is the neural wiring in the brain. At the neuronal
level, indeed, each synapse and axon has a specific direction for the
flow of electric and chemical signal. A single voxel, which is the
maximal spatial resolution of a fMRI scan, contains about $10^4$
neurons. The information flow between two voxels can be thought as
resulting from the average flow of chemical and electric signals
between all the neurons in these voxels.

To date, Granger causality \cite{granger} is a useful tool to
statistically test probabilistic causal and directional relations
between two temporal variables and since its introduction in 1969, it
has been applied in several disciplines, ranging from finance to
neuroscience and biology.  In this section we re-construct the same
brain networks induced by LTP for the rodent brain made of the active
brain areas during fMRI scans, i.e. HC, NAc and PFC, as discussed in
the main text and, in addition, we use Granger causality
\cite{granger} to infer probabilistic directions of the network's
links. We first start from the undirected network as discussed in
Section \ref{sec:brain_net} for each one of the six animals. For each
connected pair of nodes in the functional network throughout an
undirected link, we infer directionality of the connection by applying
Granger causality to the BOLD signal of the pair of voxels. We use a
confidence level $\alpha = 0.01$ and a lapse $t_l = 1$-step in the
scanning time, which correspond to $2$ seconds, this is the minimum
temporal resolution available from the fMRI in use.

Given two voxels $i$ and $j$, from their time series, we test the
hypothesis $i$ Granger-causes $j$ and, if the hypothesis is accepted,
we assign a link $i \to j$. We then test the opposite hypothesis: $j$
Granger-causes $i$. If both hypothesis are accepted we add no
directionality to the link $i - j$, the same in the case when none of
the two hypothesis is accepted. Therefore, directionality is assigned
when either $i$ Granger-causes $j$ ($i \to j$) or $j$ Granger-causes
$i$ ($j \to i$).

To find which are the influencers, i.e. the integrators, in this
directional network we develop an heuristic version of the Collective
Influence (CI) algorithm, presented in Eq. \eqref{eq:CIeq} in the SI,
which accounts for link directionality. Once the network is directed,
each node has a given in-degree ($k_i^{\rm{in}}$) and out-degree
($k_i^{\rm{out}}$) and undirected links contribute to both of them. A
natural generalization of the CI algorithm to the directed case is
then the following:
\begin{equation}
  {\rm CI^{\rm{DIR}}}_{\ell}(i) = (k_i^*-1)
\sum_{j\in\partial{\rm Ball^*}(i,\ell)} (k_j^*-1) +
  \sum_{ \substack{j\in\mathcal{F}^*(i)\ : \\ k_j^{* \rm in-inter}\ = 1}} (k_j^*-1)   
\sum_{m\in\partial{\rm Ball^*}(j,\ell)} (k_m^*-1) .
\label{eq:CIdir}
\end{equation} 
Where here, slightly differently from the undirected case, $k_i^* =
k_i^{\rm{in}} + k_i^{\rm{out}}$ is the total degree of node $i$, with
$k_i^{\rm{in}} \equiv k_i^{\rm{in-intra}} + k_i^{\rm{in-inter}}$ that
accounts for: the total in-links coming from nodes in the same network
as $i$ ($k_i^{\rm in-intra}$); and in-links coming from nodes
belonging to a different network than $i$ ($k_i^{\rm
  in-inter}$). Analogously, $k_i^{\rm{out}} \equiv\ k_i^{\rm
  out-intra} + k_i^{\rm out-inter}$, with $k_i^{\rm out-intra}$ and
$k_i^{\rm out-inter}$ having a similar meaning but for the out-degree
of node $i$. Diversely from Eq. \eqref{eq:CIeq}, the symbol
$\partial{\rm Ball^*}(i,\ell)$ indicates the \emph{directed} sphere of
influence of node $i$: this is the sphere of influence that can be
reached with a directed path starting at node $i$. Whereas $j \in
\mathcal{F}^*(i): k_j^{* \rm in-inter}\ = 1$ instead indicates the set
of nodes connected to $i$ through a \emph{directed} interlink and
which have no more interlinks with any of the other nodes in the
network.

\begin{figure}[!t]
\includegraphics[width=8.4cm]{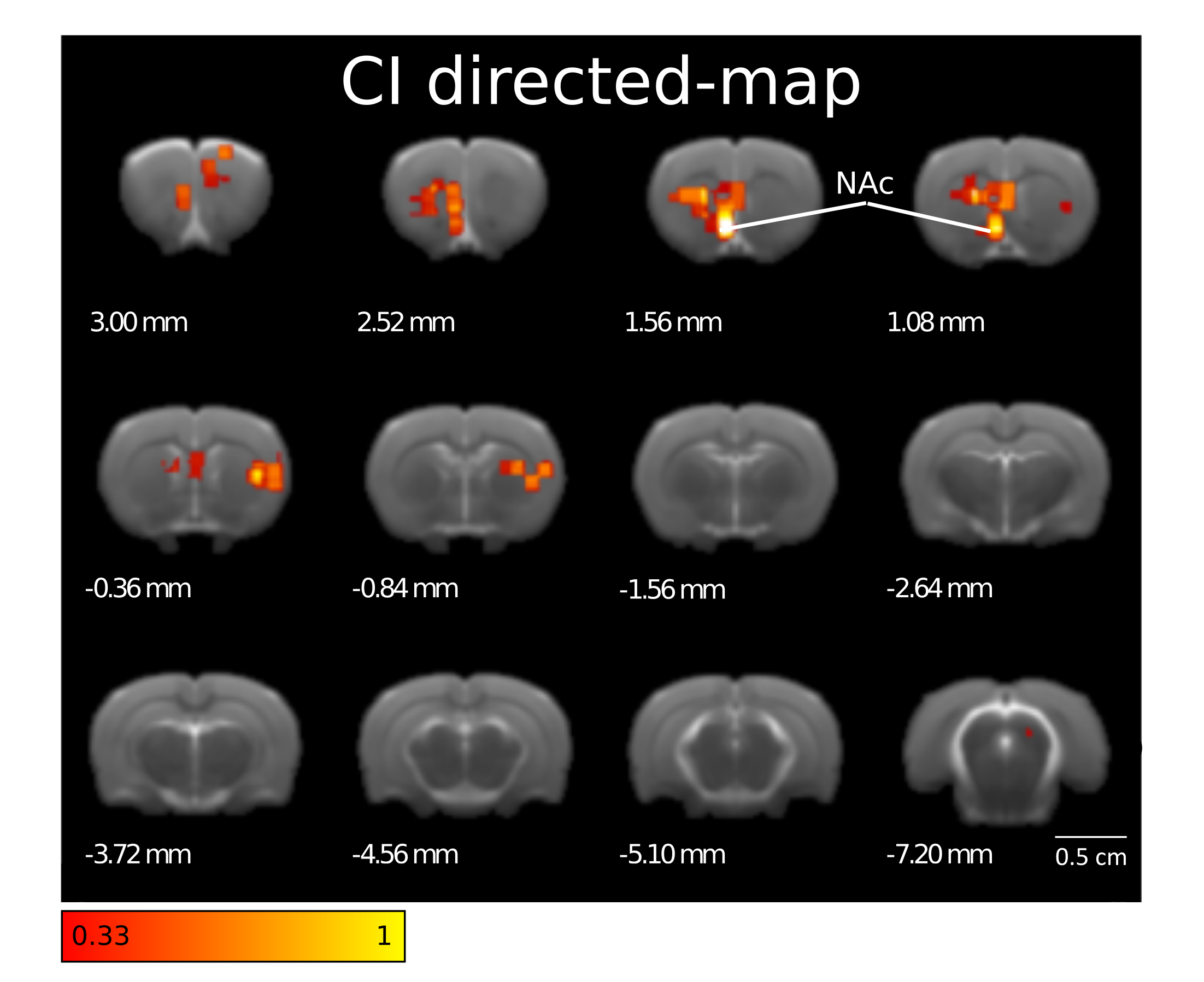}
\caption{ Average ($n=6$) CI-directed map. The map indicates the CI-score (Eq. \eqref{eq:CIdir}) of each voxel averaged over six animals for the case of directed brain network. The Nucleus Accumbens appears as the area with the highest CI averaged score and, therefore, it is identified as the main area responsible for integration.}
\label{fig:CI_dir_av}
\end{figure}

To identify the influencers of the directed brain network, for each
rat, we compute the directed CI-score according to
Eq. \eqref{eq:CIdir}, in analogy with the undirected case, for each
node in the brain network. For each animal, we then rank the nodes
from high to low score and we then compute an average CI-directed map
similarly to what described in Supplementary Note
\ref{maps}. Results are shown in Supplementary
Fig. \ref{fig:CI_dir_av}, to be compared with results for the
undirected network discussed in the main text and illustrated in
Fig. \ref{fig:fig2}h. Despite the fact that the networks are directed
in this case, the nucleus accumbens still results to be the brain area
with the highest CI-directed score and so, according to our theory,
the main brain areas responsible for integration.

\section{{Degree analysis of nodes responsible for integration}}\label{sec:degree}

\begin{figure}[t!]
\includegraphics[width=8cm]{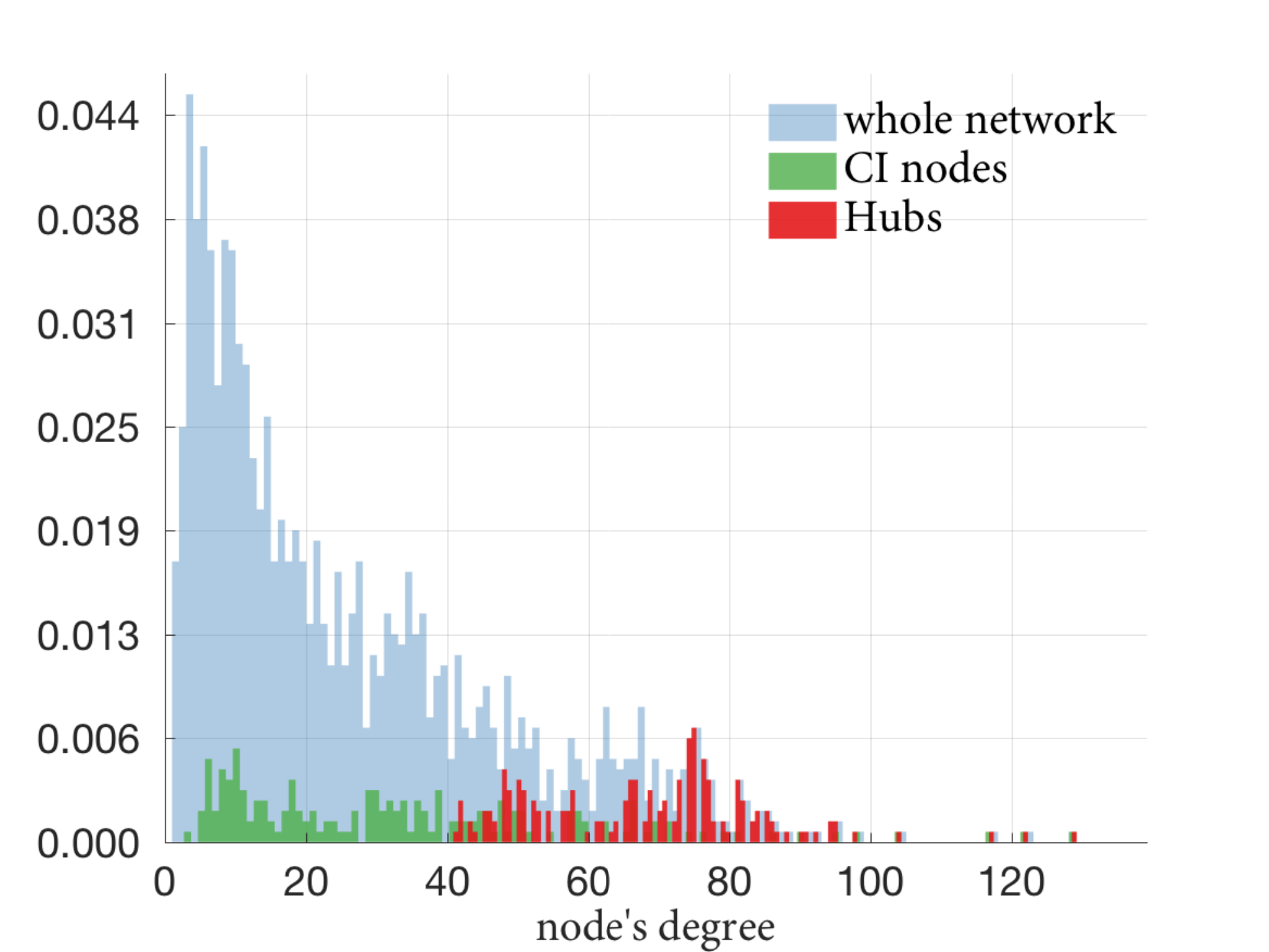}
\caption{ Degree distribution across animals ($n=6$) of: (blue bins) the whole brain network; (green bins) top 30 CI nodes; (red bins) top 30 hub nodes. The figure shows that the top CI nodes, which our theory identify as responsible for brain integration, are comparatively of lower degree than hub nodes in the same networks.}
\label{fig:degree}
\end{figure}

In this section we present a study of the degree statistics for the
top CI nodes in each rodent brain network and of the top hub nodes in
the same network. In particular, for each rat brain network, we
identify the top 30 CI nodes according to equation \eqref{eq:CIeq} and
we then determine the degree of each one of these nodes in their
relative network. Analogously, for each rat, we also identify the top
30 hub nodes by using a high-degree algorithm and then determine their
degree. We choose the first 30 nodes because, across animals, this is
the max number of CI nodes which can be removed before the network is
completely dismantled and so, the max number of nodes which can be
used to compare the CI and hub degree statistics. For completeness, we
also compute the degree statistics of all rodent brain networks by
identifying the degree of each node in the network. In Supplementary
Fig. \ref{fig:degree} we report the corresponding degree distributions
obtained from the above analyses. This figure illustrates that high CI
nodes, i.e. nodes that we find responsible for integration within our
theory, are comparatively of lower degree than hubs in the brain
network.

\section{{Pharmacogenetic (Dreadd) experiment}}
\label{si-pharmacogenetics}
 
The fundamental goal of this experiment is to use Designer Receptors
Exclusively Activated by Designer Drugs (DREADDs) technology
\cite{armbruster,roth2016} to specifically inactivate nodes in the
shell part of the nucleus accumbens (NAc), the contralateral hippocampus 
(cHC), the anterior part of the prefrontal cortex (PFC), and the
somatosensory cortex (S1) and study, using fMRI and
optimal percolation analysis, its impact on the functional
architecture in the memory network induced by LTP.  
More specifically, with the aid of adenoviral vectors, we directed the expression of a
Gi-DREADD (hM4Di) protein into the target regions which, under intra-peritoneal
administration of the otherwise inert ligand clozapine-N-oxide (CNO),
activates the receptor inducing neuronal silencing and blocking those regions output. 
Details are provided below.

\subsection*{Subjects}

A total of 15 Sprague-Dawley male rats, weighing between 260-280 g,
were used in this experiment. From these, three animals were not
considered in the analysis due to absence or poor DREADD expression in
the post-mortem validation. As before, animals were purchased from
Janvier Labs (France) and maintained under a 12/12 h light/dark cycle
(lights on 07:00-19:00 h) at room temperature (22$\pm$2 C). Food and
water were provided ad libitum.  Rats were housed in groups (4 animals
per cage) and adapted to these conditions for at least 7 days before
any manipulation.

\subsection*{Viral constructs and injection procedures}

A mixture of two viruses is used to express hM4Di in the NAc. The
first virus (AAV5-hSyn-GFP-Cre) drives the expression of Cre under the
control of Synapsin (hSyn) in neurons and provides amplification of
the Cre-dependent DREADD construct. The second virus
(AAV5-hSyn-DIO-hM4D(Gi)-mCherry) expresses the inhibitory DREADD in
Cre positive neurons. Both viruses are mixed 1:1 and 0.25 $\mu$L are
injected stereotaxically in the shell portion of the NAc. For this,
isoflurane anesthetized animals (4 $\%$ induction and 2.5 $\%$ for
maintenance in 0.8 L/min $O_2$) are fixed in an stereotaxic frame, as
described above, and bilateral craniotomies opened over the NAc 
(from bregma, AP 2.5 mm, ML 1.3 mm, and DV 7 mm), 
the PFC (from bregma AP -3.2 mm, ML 0.5
mm, and DV 2.0 and 3.8 mm), the contralateral Hippocampus 
(from bregma, AP -3.5 mm, ML 2.6 mm, and DV 3.2 mm), and the 
somatosensory cortex S1 (from bregma, AP 0.8 mm, ML 3.6
mm, and DV 1.4 mm). 
Injections are performed using silica cannula (GC22-20, 22 gauge
internal cannula, WPI, USA) coupled to an infusion pump (SP200IZ
Syringe pump, WPI, USA) through polyethylene tubing. The cannula is
lowered slowly in the tissue to the final stereotaxic coordinate,
stays in place 10 min before infusion starts, and 10 min more before
retraction. Retraction is done slowly to prevent sucking the injected
solution. At the end of the procedure, both craniotomies are covered
with small amounts of bone cement (Palacos, Heraheus Medical GmbH,
Germany), and the skin sutured.  After the surgery animals receive
analgesics (buprenorphine 0.3 mg, Buprex, Reckitt Benckiser
Healthcare, UK) and antibiotics (enrofloxacine 3 $\%$, Syvaquinol 25,
Syva, Spain) during 3-5 days.

\subsection*{DREADD fMRI procedures}

We wait 4 to 6 weeks after the injection of the viruses to allow
proper expression of the DREADD proteins in the NAc neurons. The
experimental procedures for electrode implantation and fMRI data
acquisition are the same as explained above for the LTP experiment. In
addition, animals in this experiment are intraperitoneally cannulated
for CNO administration inside the magnet. After baseline fMRI
acquisition is completed (corresponding to the PRE-LTP, PRE-CNO
condition), CNO is administered i.p (1 mg/Kg, 10 mL/Kg) and 30 min
later a first set of functional images is acquired during low
frequency stimulation (PRE-LTP, POST-CNO condition).  After that, and
still under the effect of CNO (which last more than 10 h,
\cite{alexander}), LTP is induced as before and 1h later a new set of
functional images is acquired (POST-LTP, POST-CNO condition).

\subsection*{Histology}

At the end of each experiment, rats are perfused intracardially with
100 mL of 1$\%$ phosphate-buffered saline (PBS) solution and 100 mL of
ice-cold 4$\%$ paraformaldehyde (PFA). Brains are kept for 24h on
4$\%$ PFA post-fixation at 4 C and cut in a fixed material vibratome
in 50 $\mu$m thick slices. Slices are then stained with
4',6-diamidino-2-phenylindole (DAPI) for photography under a
fluorescence microscope. Expression of hM4Di in the NAc is validated
by GFP fluorescence in the neurons.

\section{Pharmacologic (TTX) inactivation experiments}
\label{sec:TTX}

In this experiment, we used an acute infusion of the voltage-dependent sodium channel blocker Tetrodotoxin (TTX), to strongly inactivate nodes in targeted regions and study, using fMRI and optimal percolation analysis, its impact on the functional architecture in the memory brain network induced by LTP.

\subsection*{Subjects}
A total of 8 Sprague-Dawley male rats, weighing between 250-300 g, were used in this experiment, 4 for Hippocampal inactivation and 3 for S1 inactivation. As before, animals were purchased from Janvier Labs (France) and maintained under a 12/12 h light/dark cycle (lights on 07:00-19:00 h) at room temperature ($22 \pm 2$ C). Food and water were provided ad libitum. Rats were housed in groups (4 animals per cage) and adapted to these conditions for at least 7 days before any manipulation.

\subsection*{Drug and injection procedures}
Urethane anesthetized animals are fixed in a stereotaxic frame, as described above, and craniotomies are opened bilaterally over the Hippocampus (from bregma, AP -3.5 mm, ML 2.6 mm, and DV 3.2 mm), or the somatosensory cortex S1 (from bregma, AP 0.8 mm, ML 3.6 mm, and DV 1.4 mm). 
Injections are performed using silica cannula (GC22-20, 22-gauge internal cannula, WPI, USA) coupled to an infusion pump (SP200IZ Syringe pump, WPI, USA) through polyethylene tubing. The cannula is lowered slowly in the tissue to the final stereotaxic coordinate, stays in place 10 min before infusion starts, and 10 min more before retraction. $0,5 \mu$L of TTX (100 $\mu$M in ACSF) are infused in the target region. Retraction is done slowly to prevent sucking the injected solution. 
Two multichannel recording electrodes are inserted in the ipsilateral and contralateral Hippocampus to account for the induced TTX inactivation and the successful induction of LTP. After TTX is infused as described above, field potentials in the contralateral Hippocampus are abolished, whilst field potentials in the ipsilateral Hippocampus remain intact (not shown). After that, LTP induction and fMRI procedures proceed as described in the main text and Supplementary Note \ref{protocol}.

\clearpage


\end{document}